\documentclass[]{aa}  
\usepackage{graphicx}
\usepackage{txfonts}
\usepackage{natbib}
\bibpunct{(}{)}{;}{a}{}{,} 

\def\lfir{$L_{\rm{FIR}}$}
\def\xeff{$\epsilon_{\rm{xeff}}$}
\def\emech{{$\dot{E}_{K}$}}

\def\ergs{erg s$^{-1}$}
\def\ergsa{erg s$^{-1}$ \AA$^{-1}$}
\def\ergsacm{erg s$^{-1}$ cm$^{-2}$ \AA$^{-1}$}
\def\ergscm{erg s$^{-1}$ cm$^{-2}$}
\def\msun{M$_\odot$}
\def\ebv{{\em E(B-V)}}

\def\lyalp{{Ly$\alpha$}}
\def\lulyalp{{$L(\rm{Ly}\alpha)$}}

\def\luv{{$L_{UV}$}}
\def\fuv{{$F_{UV}$}}

\def\ewhb{{$EW(\rm{H}\beta)$}}

\def\hal{{$\rm{H}\alpha$}}
\def\hb{{$\rm{H}\beta$}}

\def\lha{{$L(\rm{H}\alpha)$}}
\def\lhb{{$L(\rm{H}\beta)$}}

\def\oxiil{{O[II]$3727$}}
\def\oxiiid{{O[III]$4959$, $5007$}}
\def\siivd{{Si~IV$1394$,~$1403$}}
\def\civd{{C~IV$1548$,~$1551$}}
\def\lyai{\lyalp$_{\rm{i}}$}
\def\lyaii{\lyalp$_{\rm{ii}}$}
\def\lyaiii{\lyalp$_{\rm{iii}}$}
\def\lyaiv{\lyalp$_{\rm{iv}}$}
\def\lyav{\lyalp$_{\rm{v}}$}

\def\ne{$n(\rm{e})$}
\def\lhalx{\lha$/L_{0.4-2.4\rm{\,keV}}$}

\begin{document}
\title{Multiwavelength analysis of the Lyman alpha emitting galaxy Haro~2: relation between the diffuse Lyman alpha and soft X-ray emissions 
}

\titlerunning{Multiwavelength analysis of starburst in Haro~2}

   \author{H. Ot\'{\i}-Floranes\inst{1,2}
          \and
          J.M. Mas-Hesse\inst{1}
          \and
          E. Jim\'{e}nez-Bail\'{o}n\inst{3}
          \and
          D. Schaerer\inst{4,5}
          \and
          M. Hayes\inst{6,5}
          \and
          G. \"{O}stlin\inst{7}
          \and
          H. Atek\inst{8}
          \and
          D. Kunth\inst{9}
}

   \offprints{J.M. Mas-Hesse}

   \institute{Centro de Astrobiolog\'{\i}a (INTA--CSIC), Departamento de Astrof\'{\i}sica, POB 78, 
	     E--28691 Villanueva de la Ca\~nada, Spain\\
             \email{otih@cab.inta-csic.es, mm@cab.inta-csic.es}
\and
	     Dpto. de F\'{\i}sica Moderna, Facultad de Ciencias, Universidad de Cantabria,
39005
	      Santander, Spain
\and
	     Instituto de Astronom\'{i}a, Universidad Nacional Aut\'{o}noma de M\'{e}xico, Apartado Postal 70-264, 04510, M\'{e}xico DF,
M\'{e}xico
\and
	     Observatoire de Gen\`{e}ve, Universit\'{e} de Gen\`{e}ve, 51 Ch. des Maillettes, 1290 Versoix, Switzerland
\and
	     CNRS, Institut de Recherche en Astrophysique et Plan\'{e}tologie, 14 avenue Edouard Belin, F-31400 Toulouse, France
\and
	     Universit\'{e} de Toulouse, UPS-OMP, IRAP, Toulouse, France
\and
	     Department of Astronomy, Oskar Klein Centre, Stockholm University, SE - 106 91 Stockholm, Sweden
\and
	     Spitzer Science Center, Caltech, Pasadena, CA 91125, USA
\and
	     Institut d'Astrophysique de Paris (UMR 7095: CNRS \& UPMC), 98 bis Bd Arago, 75014 Paris, France
             }

   \authorrunning{Ot\'{\i}-Floranes et al.}
   
   \date{Received 30 March 2012; Accepted 30 July 2012}
 
\abstract
{\lyalp\ emission is commonly used as star formation tracer in cosmological
studies. Nevertheless, resonant scattering strongly affects the resulting
luminosity, leading to variable and unpredictable escape fractions in different
objects.}
{To understand how the \lyalp\ escape fraction depends on the properties of the
star-forming regions, we need high spatial resolution multiwavelength studies of
nearby \lyalp\ emitters, like Haro~2.}
{We study the \lyalp\ emission of Haro~2 in connection with the properties of
the young stellar population, the characteristics of the interstellar medium,
the distribution and intensity of the Balmer emission lines and the properties
of the X-ray emission. We have used {{\em HST}} FOC UV, WFPC-2 optical and
NICMOS NIR images, as well as STIS spectral images along the major and minor
axes of Haro~2 to characterize the \lyalp\ emission and analyze the properties
of the stellar population. WFPC-2 \hal\ image and ground-based spectroscopy
allow us to study the Balmer emission lines. Finally, {\em Chandra}/ACIS X-ray
images provide resolved distribution of the X-ray emission at various energy
bands. The observational data are compared with the predictions from
evolutionary synthesis models to constrain the properties of the star formation
episode.}
{The UV, \hal\ and far infrared luminosities of the Haro~2 nuclear starburst are
well reproduced assuming a young stellar population with ages $\sim$$3.5-5.0$
Myr, affected by differential intestellar extinctions. Whereas the diffuse soft
X-ray emission observed is attributed to gas heated by the mechanical energy
released by the starburst, an unresolved component (likely an UltraLuminous
X-ray source) accounts for the hard X-ray emission. Both compact and diffuse
\lyalp\ emission components are observed along both axes. \lyalp\ is spatially
decoupled from Balmer lines emission, Balmer decrement and UV continuum.
However, the diffuse \lyalp\ component is spatially correlated with the diffuse
soft X-ray emission. Moreover, unlike the compact \lyalp\ emission, diffuse
\lyalp\ shows luminosities larger than predicted from \hal, assuming case B
recombination and considering the dust extinction as derived from \hal$/$\hb.}
{The presence of outflows and dust abundance strongly affects the \lyalp\
emission closely associated to the massive stellar clusters, leading to even a
range of escape fractions at different locations within the same starburst. On
the other hand, we propose that the diffuse \lyalp\ emission originates in gas
ionized by the hot plasma responsible for the soft X-ray radiation, as suggested
by their spatial correlation and by the measured \lhalx\ ratios. Calibration of
\lyalp\ as star formation rate tracer should therefore include both effects
(destruction vs. enhancement) to avoid biases in the study of galaxies at
cosmological distances.}


   \maketitle

\section{Introduction}

The spectral energy distribution (SED) of galaxies experiencing intense and relatively
short episodes of massive star formation ({\em starbursts}) is dominated in the UV
by the continuum of the young, most massive stars. The ionizing radiation they
produce interacts with the surrounding gas to produce nebular emission lines through
recombination processes. The mechanical energy released during the evolution of these
stars, in the form of stellar winds and supernova explosions, heats the medium, producing
X-ray emission and collisionally excited emission lines. While \hal\ is
the most prominent hydrogen nebular line in the optical, simple nebular models predict
\lyalp\ to be $\sim$$9$ times more intense than \hal\ when considering typical nebular
parameters for temperature and electronic density \citep{Osterbrock89}. Nevertheless,
observations over the last decades have shown that the value of the \lulyalp$/$\lha\ may
indeed be very different than expected, and that the presence, intensity, spectral shape
and spatial profile of \lyalp\ line do depend on many factors (see \citet{MasHesse03}, \citet{Atek09}, \citet{Hayes10},
\citet{Ostlin09} and references therein). \lyalp\ is a resonant line and this causes
\lyalp\ photons to be absorbed and reemited by neutral hydrogen atoms within the 
gas many times before they can escape. The increase in the path of the photons at \lyalp\
wavelength due to this resonant effect implies that very little dust can completely
destroy the photons around $1216$ \AA, yielding a damped \lyalp\ absorption \citep{Chen94,Atek09}. However, as
explained by \citet{MasHesse03} and \citet{Verhamme06}, when the mechanical energy
injected by the starburst into the interstellar medium (ISM) accelerates the surrounding
neutral gas, forming an expanding superbubble, only the blue wing of the emission line is
scattered. As a result, \lyalp\ emission can escape from the region, but will show a
characteristic P~Cyg emission--absorption profile. \citet{MasHesse03} showed different
cases in a sample of blue compact galaxies in the Local Universe, ranging from damped
absorption to emission with and without P~Cyg profiles, depending on the kinematical and
ionization properties of the medium surrounding the star-forming regions.
\citet{Verhamme06} computed the different profiles expected applying radiation transfer
models to different scenarios, including the presence/absence of accelerated gas, dust,
energy shift of the photons by the scattering process,... 

\citet{Tenorio99} proposed a sequence of \lyalp\ profiles following the different
evolutionary phases of the starburst, assuming the acceleration of the neutral gas by the
ongoing starburst and the formation of expanding shells to be the driver of the \lyalp\
line visibility. Since the acceleration of the surrounding interstellar medium is
responsible for both the generation of X-ray emission, by the process of gas heating
\citep{Strickland99}, and for the escape of the \lyalp\ emission line, by opening
kinematical 'holes' in the neutral gas, we expect some intrinsic correlation between both
phenomena. In order to study this possible correlation we embarked in a
project to study the X-ray properties of \lyalp\ emitters in the Local Universe, starting
with Haro~2.

\begin{figure*}
\centering
\includegraphics[width=18cm,bb=0 0 1271 609
dpi,clip=true]{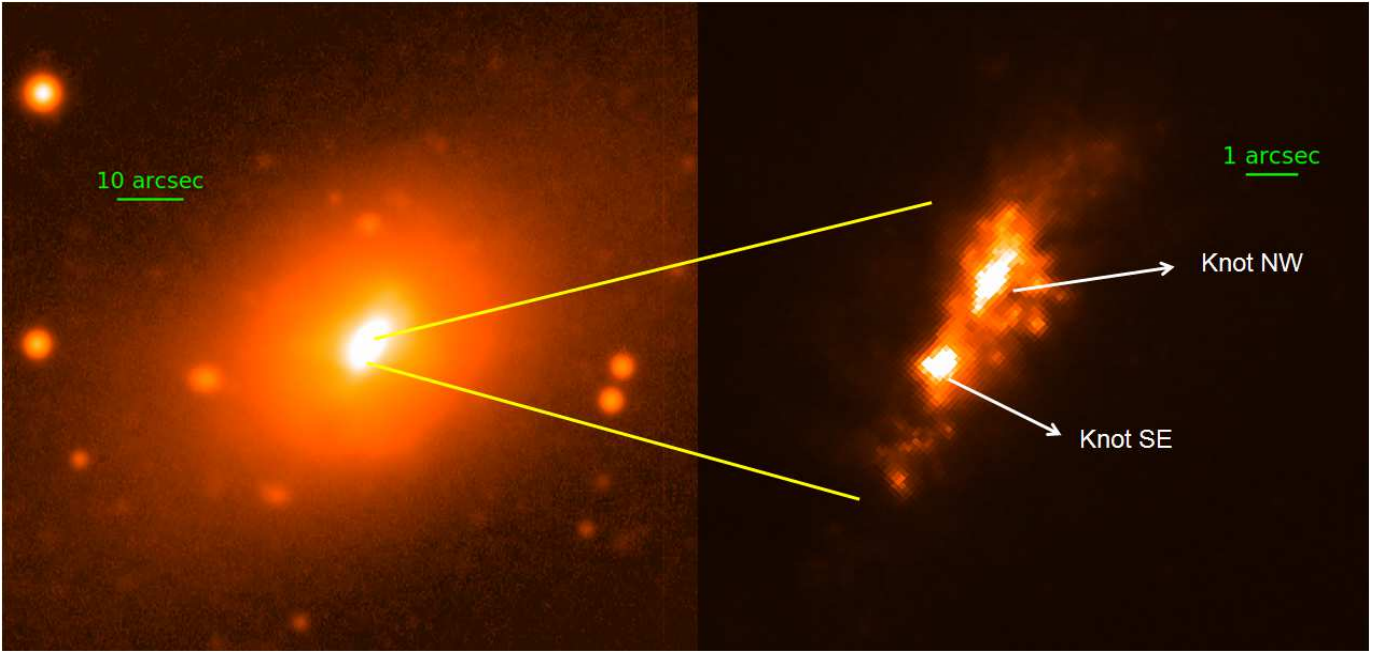}
\caption{Left: image of the whole galaxy Haro~2 from {\em NOT}/ALFOSC in the $V$-band. Right:
image of the UV-continuum of the nucleus of Haro~2 obtained with {\em HST}/WFPC-2 with
filter F336W where knots SE and NW are labelled.}
\label{figharo}
\end{figure*}

\begin{table*}
\caption{Haro~2 coordinates, distance, scale, Galactic HI, color excess
towards the source, and oxygen abundance.}
\label{harotable}
\centering
\begin{tabular}{cccccccc}
\hline\hline\\
\multicolumn{1}{c}{R. A.} & \multicolumn{1}{c}{Decl.} & Redshift$^{\mathrm{a}}$ &
\multicolumn{1}{c}{Distance$^{\mathrm{a}}$} & \multicolumn{1}{c}{Scale$^{\mathrm{a}}$} &
\multicolumn{1}{c}{$N($HI$)^{\mathrm{b}}$} & \ebv$_{\rm{Gal}}$ &
$12+\rm{log}(\rm{O}/\rm{H})^{\mathrm{c}}$ \\
(J2000.0) & (J2000.0) & & (Mpc) & (pc arcsec$^{-1}$) & (cm$^{-2}$) & & \\
\hline\\
$10$ $32$ $31.9$ & $+54$ $24$ $03.7$ & $0.004769$ & $20.5$ & $100$ & $6.3\times10^{19}$ & $0.012$ &
$8.5$ \\
\hline
\end{tabular}
\begin{list}{}{}
\item[$^{\mathrm{a}}$] From NED (the NASA/IPAC Extragalactic Database).
\item[$^{\mathrm{b}}$] Value from \citet{Lequeux95}.
\item[$^{\mathrm{c}}$] Value from \citet{Davidge89}.
\end{list}
\end{table*}

Haro~2, also known as Mrk 33, Arp 233 and UGC 5720, is a metal-rich blue compact dwarf
galaxy \citep{Davidge89}, experiencing an intense star-forming episode with an age of
$\sim$$5$ Myr \citep{MasHesse99b}. Star formation is located in the nucleus of the galaxy
where it can be resolved into individual knots (Fig.~\ref{figharo}). \citet{Chandar04} combined the {\em
HST}/STIS spectra of six different clusters within the starbursting nucleus, whereas
\citet{Mendez00} could only resolve three knots in an \hal\ image obtained with the HIRAC
camera of the Nordic Optical Telescope. On the other hand, \citet{Summers01} detected two
knots in a B-band image from the Jacobus Kapteyn Telescope. Besides the young population
produced in the current star-formation episode \citet{Fanelli88} argued for the existence of
an older stellar population given the spectral features of A-type stars found by the
International Ultraviolet Explorer (IUE). Based on the data by \citet{Fanelli88},
\citet{Summers01} decomposed the star formation history of Haro~2 into a current starburst
of $\sim$$5$ Myr, a previous burst of $\sim$$20$ Myr and an earlier episode having ocurred
$\sim$$500$ Myr ago. Despite being a metal rich, rather dusty galaxy, \citet{Lequeux95} found
that Haro~2 is an intense \lyalp-emitter. The line shows a clear P~Cyg profile apparently
originated by a neutral superbubble expanding at $\sim$$200$ km s$^{-1}$, energized by the
central starburst. \citet{MasHesse03} obtained a high-resolution spectral image with {\em
HST}/STIS in the \lyalp\ range along the minor axis, showing the P~Cyg profile and the
spatial decoupling of the continuum and the \lyalp\ emission. \citet{Summers01} performed
a study of the X-ray radiation of Haro~2 based on {\em ROSAT}/HRI data and proved that the
mechanical energy released by the nuclear starburst was indeed powering the diffuse, soft
X-ray emission. These authors quantified in $2$\% the fraction of the mechanical energy
injected by the burst into the medium ending up being emitted as X-rays. \citet{Legrand97}
identified the large \hal\ shell at the edges of the expanding superbubble, decoupled from
the rotation velocity of the galaxy.

The bolometric luminosity of Haro~2 is well represented by its infrared luminosity
\lfir$=1.4\times10^{43}$ \ergs, which overlaps with the $L_{bol}$ low-end in the luminosity function of Lyman Alpha
Emitting Galaxies (LAEs) at $z=3.1$ \citep{Gronwall07}. Haro~2 also shows an UV
continuum luminosity similar to that of the weakest Lyman Break Galaxies (LBGs) in the
redshift range $z=2.7-5$ \citep{Tapken07}. This fact, together with its high
\lyalp-intensity and P~Cyg spectral profile, makes of Haro~2 a good local prototype of
high-$z$ \lyalp-emitters whose star formation activity is usually derived from both
\lyalp\ luminosity, UV continuum and/or X-ray emission. Understanding the processes
determining these emissions in a closeby galaxy which can be studied in high detail is of
paramount importance to quantify the star formation activity in distant galaxies. The
basic properties of Haro~2 are summarized in Table~\ref{harotable}.  

In this paper we have analyzed observations obtained by {\em HST} with STIS, NICMOS and
WFPC-2, and by {\em Chandra} in X-rays, as well as data obtained from ground-based
observatories, to fully characterize the properties of the star formation episodes taking
place in the nucleus of Haro~2 and of its associated X-ray and \lyalp\ emissions. In Sect.
\ref{obsdata} we describe the {\em HST}, {\em Chandra} and ground-based observations. In Sect.
\ref{results} we present the results obtained from the analysis of the observational data,
which we discuss later in Sect.~\ref{discussion}. Sect.~\ref{conclusions} contains the
conclusions of the study.

\begin{figure*}
\centering 
\includegraphics[width=9.0cm,bb=80 0 425 322
dpi,clip=true]{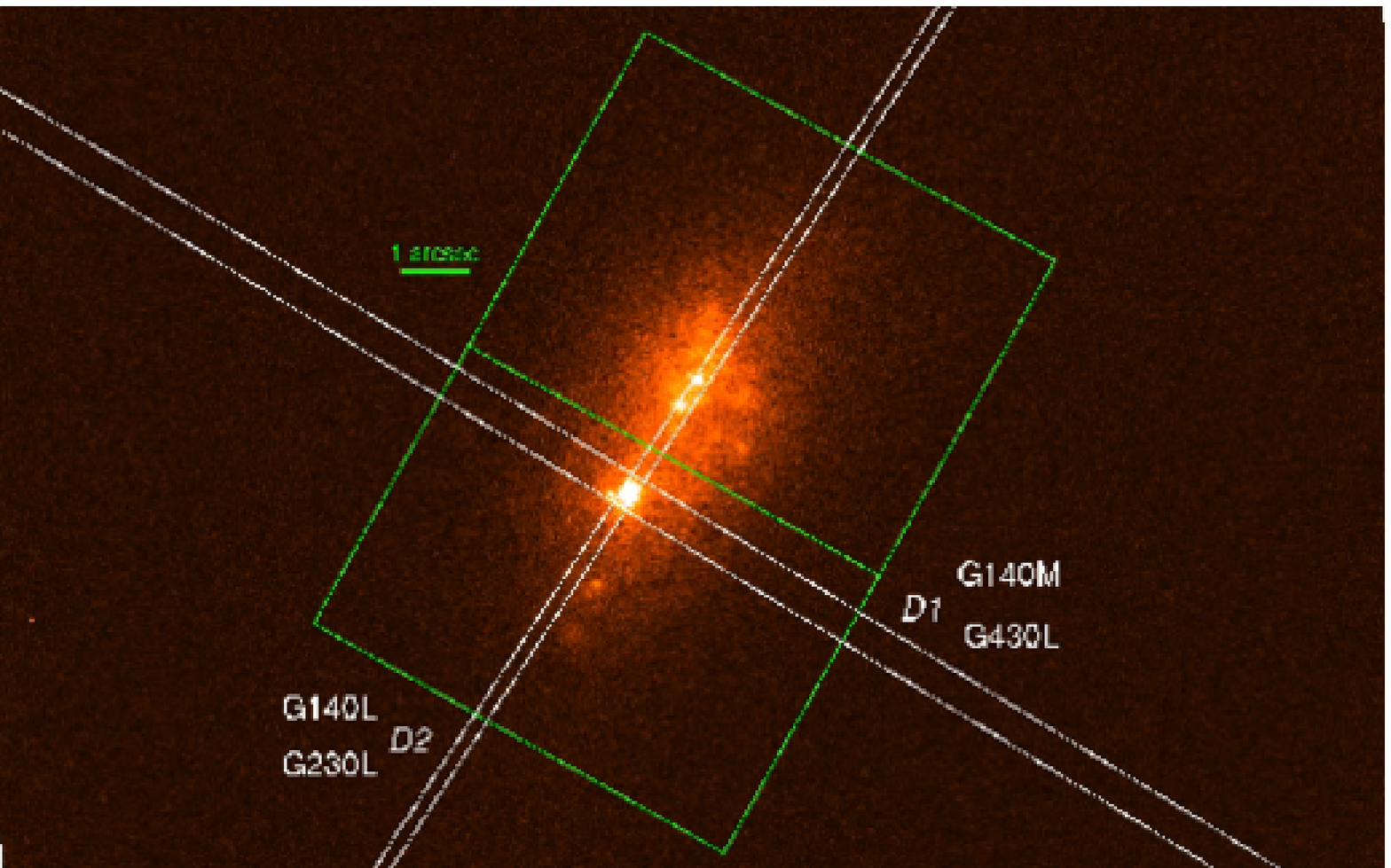}
\hspace{0.2 cm}
\includegraphics[width=9.0cm,bb=80 20 425 342
dpi,clip=true]{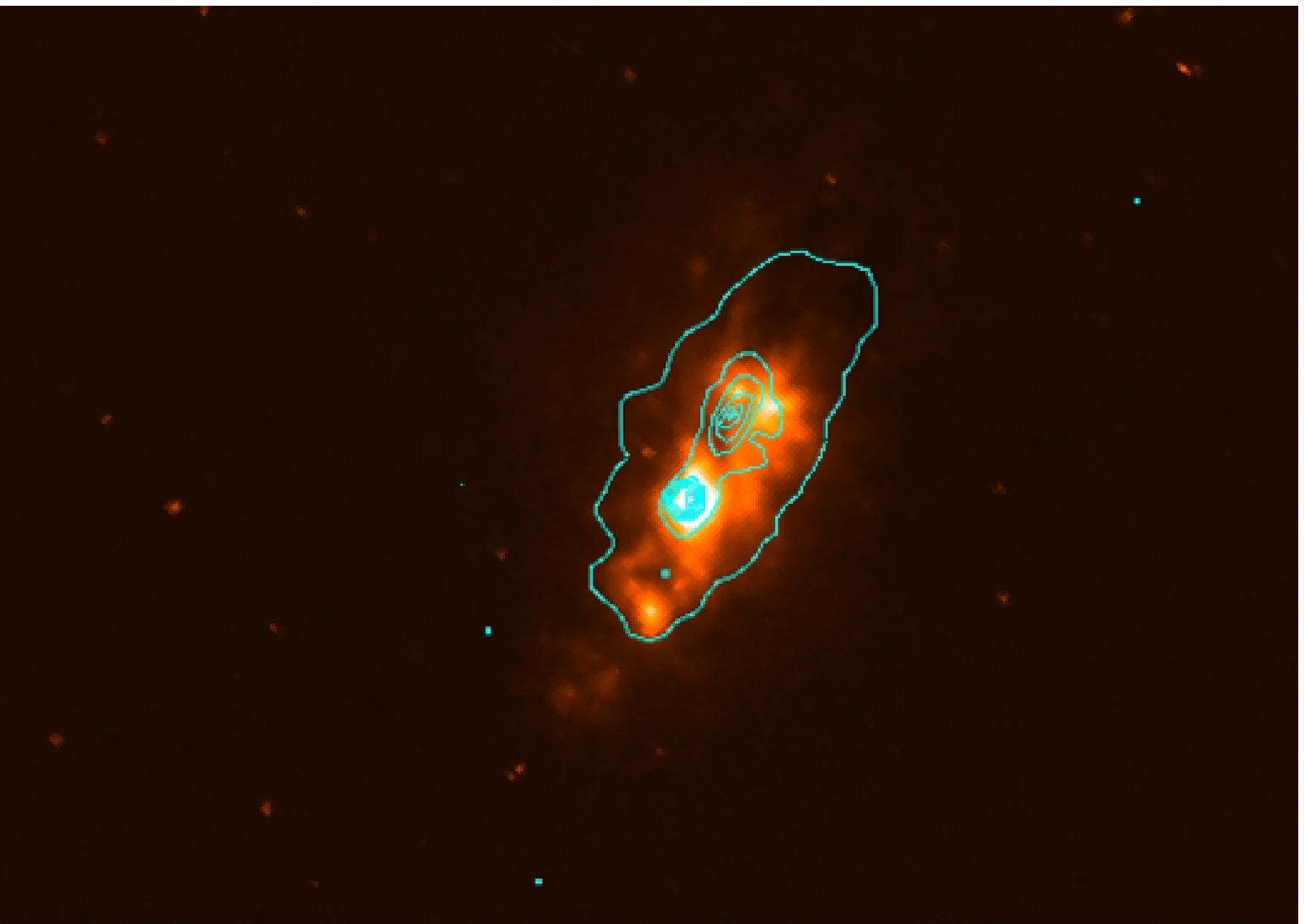}
\caption{Left: Ultraviolet {\em HST}/FOC image of Haro~2, together with the locations of
the {\em HST}/STIS slits D1 (gratings G140M and G430L) \citep{MasHesse03}, and D2 (G140L
and G230L) \citep{Chandar04}, and the boxes used to extract the UV flux from each knot.
The width of the slits is represented at real scale. North is up and East is left. Right:
\hal\ image obtained with {\em HST}/WFPC2, with the UV continuum contours superimposed.}
\label{figuvopt-halpha}
\end{figure*}

Throughout this paper we will consider $20.5$ Mpc as the distance to Haro~2, which
yields a projected scale of $\sim$$100$ pc arcsec$^{-1}$. We want nevertheless to remark
that the distance to Haro~2 could be larger, when compared to IZw18, whose estimated
distance has been increased by a factor of almost $2$ to $19.5$ Mpc by the analysis of Colour
Magnitude Diagrams and Cepheid variables. The absolute values concerning luminosities,
masses and so on should be taken with a word of caution, although they do not affect our
conclusions.

\section{Observational data} 

\label{obsdata}

\subsection{{\em HST} observations: ultraviolet, optical and near infrared}

We have used  observations of Haro~2 obtained with STIS, WFPC-2 and NICMOS, in the UV,
optical and near infrared ranges, obtained from the HST Archive at STScI. In this section
we describe the data and the processing done.

\subsubsection{{\em HST} images}

The {\em HST} imaging data sets used are listed in Table~\ref{imagetable}. There
is a single UV image of Haro~2, obtained with FOC in 1993, before COSTAR was
installed. The observation was performed with the F220W filter. The far-UV image
covers $\sim$$21\arcsec \times 23\arcsec$ with plate scale of
$\sim$$0.02\arcsec$ pixel$^{-1}$, and is centered on $\lambda=2328$ \AA. The FOC
image is shown in Fig.~\ref{figuvopt-halpha}, together with the STIS slits
directions perpendicular to the major axis (direction D1) and parallel to it
(direction D2). Since UV emission is dominated in star-forming regions by
massive stars, with this FOC image we can study the UV-morphology of the
starburts located in the nucleus of the galaxy, as well as calculate the UV-flux
at $\lambda=2328$ \AA\ (\fuv, hereinafter) emitted by each knot and by the total
source as a whole. To derive the \fuv\ flux of each knot we defined two similar
and non-overlapping boxes, as indicated in Fig.~\ref{figuvopt-halpha}. The
boundaries of these boxes are rather arbitrary, since we can not discriminate
beween both components, but the flux inside them should be dominated by the
bright stellar clusters within knots NW and SE. The sizes of the boxes are
$7.4\arcsec \times 5.5\arcsec$ (knot NW) and $7.4\arcsec \times 4.8\arcsec$
(knot SE). We have checked that the brightest knots account indeed for
$70-80$\% of the total UV flux measured on each box, dominating so their
integrated spectral properties. As we will discuss later, while the properties
of the starburst will be therefore dominated by the massive stars in the bright
knots, we have used the total flux within the boxes to derive the mass
normalization when applying evolutionary synthesis models.

\begin{figure*}
\centering
\includegraphics[width=16cm,clip=true]{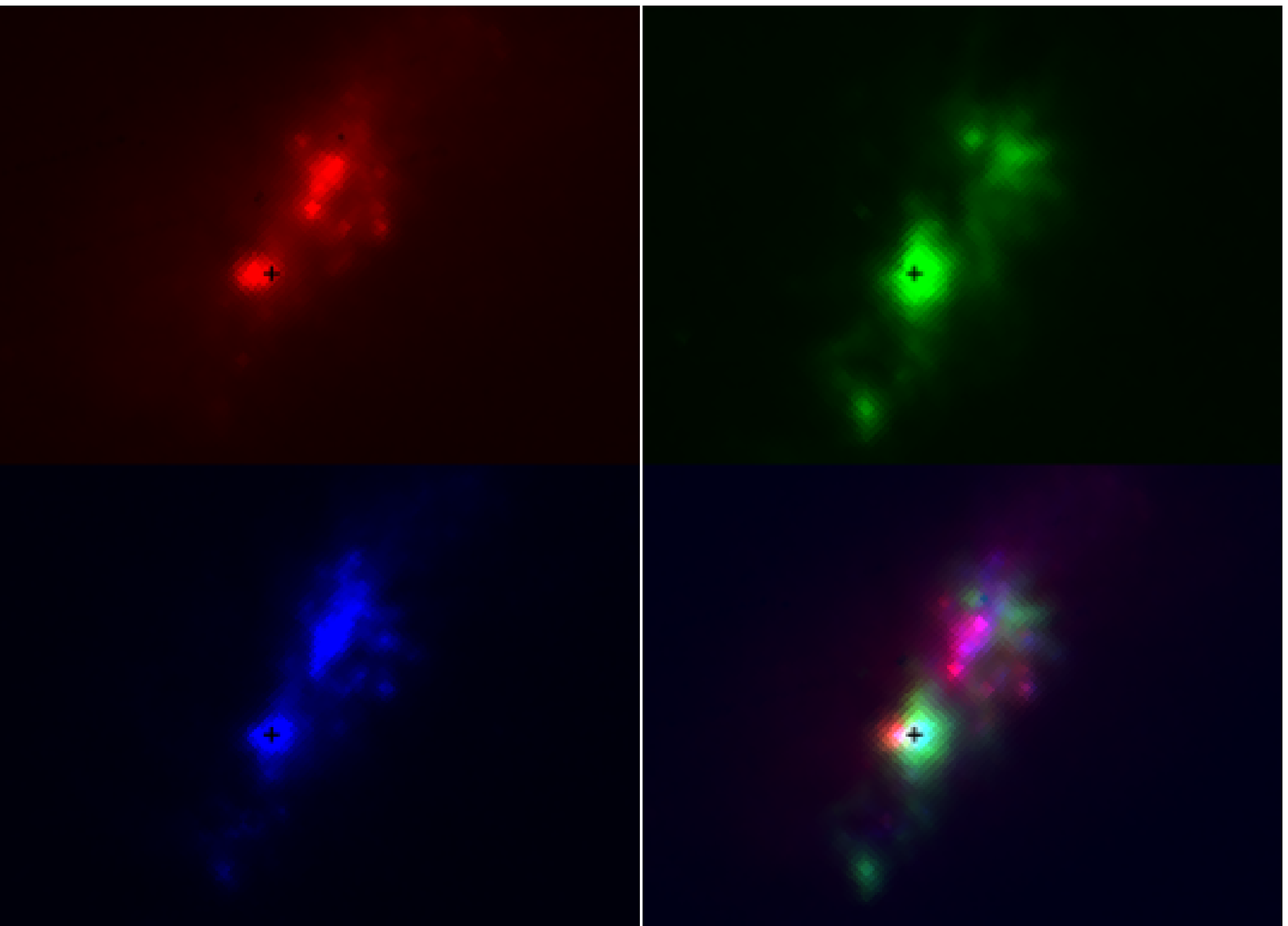}
\caption{Multiwavelength images of Haro~2: NICMOS F160W near-infrared continuum (red), WFPC-2 F336W UV
continuum (blue) and \hal\ emission (green). Images are $10$ arcsec wide. The black cross
marks the center of the bluest stellar cluster in knot SE. Note the red stellar population
to the East of knot SE, and disperse within knot NW. North is up and East is left.}
\label{fignicwf}
\end{figure*}

Background on the FOC image was estimated from the most separated regions away
from the source, averaging on boxes of $3\times3$, $5\times5$ and $10\times10$
pixel$^{2}$. The UV-background was estimated to be  $\sim$$0.75$ counts
pixel$^{-1}$.  In order to work with an image free from background
contamination, we subtracted this average background from all pixels of the
image. After that, the intensity was multiplied by {\tt PHOTFLAM} and divided by
the total exposure time {\tt EXPTIME} in order to work with fluxes in units of \ergsacm.

We studied the effect of FOC point spread function (PSF) since the image was obtained
prior to {\em COSTAR} deployment. We downloaded a pre-{\em COSTAR} PSF for an F/96 F220W
observation performed five months earlier from the instrument website\footnote{{\rm
http://www.stsci.edu/hst/foc/calibration/f96\_nov1992.html}} and extracted the enclosed
point source flux within the boundaries of the boxes we have considered. It was found that
for the sizes and locations of the regions considered and assuming the knots are
point-like sources, a $\sim$$85$\% of the total radiation is included within the integration
area, hence leading to an underestimation in the total \fuv\ of less than $\sim$$15$\%.

Haro~2 has been observed with {\em HST}/WFPC2 in the UV-optical range with broad-band filters
F336W, F439W, F555W and F814W, providing images of the continuum emission, and with the narrow
filter F658N, which is dominated by \hal. In order to isolate the \hal-emitting
regions, continuum was subtracted from the F658N image using the F814W one, assuming it
represents well the continuum at \hal. After downloading the data from the {\em HST} archive,
drizzled files were combined and aligned for filters F814W and F658N. Then, the resulting
images were multiplied by {\tt PHOTFLAM} in order to convert counts into flux values.
Finally, the F814W image was multiplied by an arbitrary constant and then subtracted from
the F658N image, using field stars as reference.

Finally, we downloaded from {\em HST} archive an image of Haro~2 obtained by NICMOS at
$1.60$, $\mu$m with broad-band filter F160W. Its size is $19.2\arcsec \times 19.2\arcsec$,
and the scale of the pixel is $0.075\arcsec$. No background subtraction on the science data
file was necessary since NICMOS reduction pipeline already performs this removal through
task {\tt cnib}. The NICMOS F160W, WFPC-2 F336W and WFPC-2 \hal\ images have been
colour-coded and combined in Fig.~\ref{fignicwf}.

\begin{table*}
\caption{Log of {\em HST}/FOC, {\em HST}/NICMOS, {\em HST}/WFPC2 and {\em Chandra} observations of
Haro~2 used in this work.}
\label{imagetable}
\centering
\begin{tabular}{lccccc}
\hline\hline\\
Instrument & Observation date & Filters & \multicolumn{1}{c}{Integration time} &
\multicolumn{1}{c}{$\lambda_{0}$} & \multicolumn{1}{c}{Plate scale} \\
&  &  & (s) & (\AA) & (arcsec pixel$^{-1}$) \\
\hline\\
{\em HST}/FOC$^{\mathrm{a}}$ & $1993$ Apr $18$ & F220W & $1197$ & $2328$ & $0.02$  \\ 
{\em HST}/NICMOS & $2006$ Sep $28$ & F160W & $912$ & $1.6\times10^{4}$ & $0.075$ \\
{\em HST}/WFPC2$^{\mathrm{b}}$ & $2009$ Mar $9$ & F336W & $3300$ & $3332$ & $0.10$ \\ 
{\em HST}/WFPC2$^{\mathrm{b}}$ & $2009$ Mar $9$ & F658N & $1800$ & $6591$ & $0.10$ \\ 
{\em HST}/WFPC2$^{\mathrm{b}}$ & $2009$ Mar $9$ & F555W & $600$ & $5304$ & $0.10$ \\ 
{\em HST}/WFPC2$^{\mathrm{b}}$ & $2009$ Mar $9$ & F814W & $600$ & $8337$ & $0.10$ \\
{\em Chandra}/ACIS$^{\mathrm{c}}$ & $2008$ Feb $2$ & - & $19194$ & $-$ & $0.49$ \\ 
\hline
\end{tabular}
\begin{list}{}{}
\item[$^{\mathrm{a}}$] Observation performed before {\em COSTAR} deployment.
\end{list}
\end{table*}

The astrometry included in the {\em World Coordinate System} ({\em WCS}) keywords in the headers
of the different images was not always consistent, with relative offsets of up to
$\sim$$1.2\arcsec$ between the NICMOS and FOC images, for example. We have taken the {\em
HST}/NICMOS header information as reference, which  matched  well the {\em Chandra}
one, and have corrected the astrometry from other images when needed.

\subsubsection{{\em HST} spectroscopy}

\citet{MasHesse03} studied spectral images obtained with the G140M and G430L grating on
STIS, with a $52\arcsec \times 0.5\arcsec$ long slit placed orthogonal to the major axis
of the nucleus. This high resolution spectral images covered the range $1200-1250$ \AA,
including the \lyalp\ line, and $2900-5700$ \AA, comprising the oxygen lines \oxiil\ and
\oxiiid, and \hb. Although the integration time on this optical spectral image was very
short, it provides the spatial profiles of these lines with high angular resolution.
Hereinafter, we will refer to the direction of this slit as D1. Furthermore,
\citet{Chandar04} analyzed two additional spectral images using the G140L and G230L
gratings with the $52\arcsec \times 0.2\arcsec$ long slit oriented along the major axis of
the starbursting region. These spectral images cover the spectral ranges $1150-1730$ \AA,
showing the \lyalp\ emission and the stellar absorption lines \siivd\ and \civd, and
$1570-3180$ \AA. This direction will be referred to as D2. Information on both UV-optical
observations is listed in Table~\ref{stistable}, and in Fig.~\ref{figuvopt-halpha} both
slit directions can be observed superimposed on the FOC image. Both slit directions are
orthogonal, and whereas \citet{MasHesse03} encloses only the SE star-forming knot, slit
from \citet{Chandar04} contains the bulk emission from both SE and NW knots. The G140M D1
and G140L D2 spectral images constitute the direct observation of the \lyalp\ morphology
and spectroscopy of Haro~2 which we analyze in this work.

\begin{table*}
\caption{Log of {\em HST}/STIS observations of Haro~2 used in this work.}
\label{stistable}
\centering
\resizebox{\textwidth}{!}{
\begin{tabular}{lcccccccc}
\hline\hline\\
Observation date & Grating & \multicolumn{1}{c}{Integration time} & \multicolumn{1}{c}{Position angle} & Slit direction &
\multicolumn{1}{c}{Slit size}  & \multicolumn{1}{c}{Wavelength interval} & \multicolumn{1}{c}{Plate scale} &
\multicolumn{1}{c}{Spectral dispersion} \\
& & (s) & (deg) & &  & (\AA) & (arcsec pixel$^{-1}$) & (\AA\ pixel$^{-1}$) \\
\hline\\
$2000$ Dec $1$ & G140M & $7900$ &  $-121.35$ & D1 & $52\arcsec\times0.5\arcsec$ & $1200-1250$ & $0.029$ & $0.053$ \\ 
$2000$ Dec $1$ & G430L & $300$ &  $-121.35$ & D1 & $52\arcsec\times0.5\arcsec$ & $2900-5700$ & $0.050$ & $2.746$ \\
$2003$ Feb $26$ & G140L & $1446$ & $146.85$ & D2 & $52\arcsec\times0.2\arcsec$ &$1150-1730$ & $0.025$ & $0.584$ \\ 
$2003$ Feb $26$ & G230L & $600$ &  $146.85$ & D2 & $52\arcsec\times0.2\arcsec$ & $1570-3180$ & $0.025$ & $1.548$ \\ 
\hline
\end{tabular}
}
\end{table*}

\begin{table*}
\caption{Log of {\em WHT}/ISIS observations of Haro~2 used in this work.}
\label{isistable}
\centering
\resizebox{\textwidth}{!}{
\begin{tabular}{lccccccccc}
\hline\hline\\
Observation date & Arm & Grating & \multicolumn{1}{c}{Integration time} & \multicolumn{1}{c}{Position angle} & Slit direction &
\multicolumn{1}{c}{Slit size}  & \multicolumn{1}{c}{Wavelength interval} & \multicolumn{1}{c}{Plate scale} &
\multicolumn{1}{c}{Spectral dispersion} \\
& & & (s) & (deg) & & & (\AA) & (arcsec pixel$^{-1}$) &
(\AA\ pixel$^{-1}$) \\
\hline\\
$2006$ Dec $25$ & Blue arm & R600B & $3600$ &  $-121.35$ & D1 & $3.3\arcmin\times1.0\arcsec$ & $3460-5280$ & $0.20$ & $0.45$ \\ 
$2006$ Dec $25$ & Red arm & R600R & $3600$ &  $-121.35$ & D1 & $3.3\arcmin\times1.0\arcsec$ & $5690-7745$ & $0.22$ & $0.49$ \\ 
$2002$ Feb $9$ & Blue arm & R300B & $3600$ &  $146.85$ & D2 & $4.0\arcmin\times1.1\arcsec$ & $3480-7020$ & $0.20$ & $0.86$ \\
\hline
\end{tabular}
}
\end{table*}

Extraction, background removal and analysis of the STIS D1 data were already discussed in
detail by \citet{MasHesse03}. Data were reduced with STSDAS procedures and flux from 100
rows above and below the stellar continuum were averaged to estimate the geocoronal
\lyalp\ emission and the detector background, being the latter negligible. Slit
correction was performed by multipliying measured counts by the STIS parameter {\tt
diff2pt}. Observational data for filters G140L and G230L in D2 were downloaded from the
{\em HST} archive already calibrated. Again, intensity values measured were multiplied by
{\tt diff2pt} in order to correct for slit losses. After that, the first step was to
select the continuum regions, deriving continuum extensions within the slit of $\sim$$60$ pc
and $\sim$$150$ pc for both knots SE and NW respectively (see Table \ref{knotstable}). The
integration of the contributions from each of the regions of the spectral images, and the
merging of the observations performed with each grating were done with {\em IRAF-STSDAS}
procedures.

We noticed that keywords {\tt RA\_APER} and {\tt DEC\_APER} in the headers of the G140M
and G430L {\em HST}/STIS file did not correspond to the actual position of the slit,
centered on knot SE, but rather referred to an intermediate position closer to knot NW.
From the \hal\ image obtained with {\em HST}/WFPC-2 data we extracted two different
spatial profiles, one assuming the position of the slit indicated by the {\em HST}/STIS
observation file and which is centered close to knot NW, and another profile for a slit
placed on knot SE. The angle of the slit was fixed to the value indicated in the file
header. Given the same nature of Balmer lines \hal\ and \hb, we compared the extracted
\hal\ profiles with the \hb\ profile from {\em HST}/STIS G430L observation, and they only
agree when the slit is assumed to be centered on knot SE. We did the same comparison between
profiles extracted from the {\em HST}/WFPC-2 F555W image and the profile of the composite
\hb$+$\oxiiid$+${\em continuum} obtained from the {\em HST}/STIS G430L observation and the
match is only possible when the slit is placed on knot SE. Therefore, we concluded that {\em
HST}/STIS G140M and G430L observations of Haro~2 are centered on knot SE and in
Fig.~\ref{figuvopt-halpha} this slit is located accordingly.

\subsection{{\em Chandra}: X-rays}

\label{obsxrays}

\begin{figure}
\centering
\includegraphics[width=8cm,bb=160 0 620 397
dpi,clip=true]{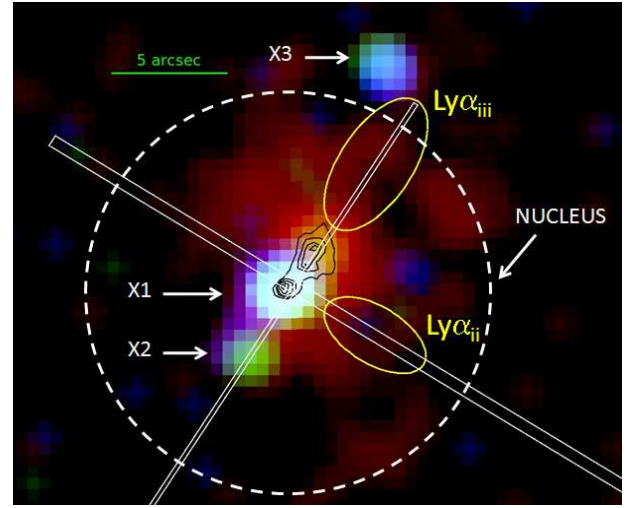} 
\caption{Color-coded {\em Chandra} X-ray image of Haro~2: soft ($0.2-1.5$
keV, red), medium ($1.5-2.5$ keV, green) and hard X-rays ($2.5-8.0$ keV, blue). The contours of the NICMOS near-infrared continuum and the position of the STIS slits have been superimposed. Main X-ray components are labeled. Source region from which X-ray spectrum was extracted is confined within the dashed circle. Regions within the yellow ellipses show diffuse \lyalp\ emission.}
\label{figx}
\end{figure}

{\em Chandra} time was allocated under proposal $09610593$ (PI: J. Miguel
Mas-Hesse) to observe Haro~2 with ACIS. Observation $9519$ with time duration of
$\sim$$20$ ks (see Table~\ref{imagetable}) was performed on February 2, 2008. {\em
Chandra}/ACIS data were reduced with {\em CIAO 4.1} following the standard
reduction procedures available on the {\em Chandra-CIAO} website, reprocessing
level 1 and 2 event files. The Ancillary Response File (ARF) and the
Redistribution Matrix File (RMF) were also calculated. In order to obtain the
X-ray spectrum of Haro~2 source counts were extracted from a circular region of
radius $\sim$$9\arcsec$ centered on the bulge emisison since it incloses most of
the X-ray radiation of the source. This region is shown in Fig.~\ref{figx}
labeled as {\em NUCLEUS}. Background was extracted from a composite of three
circular regions with radius $\sim$$46\arcsec$, $56\arcsec$, $58\arcsec$ from the
area surrounding the galaxy and not containing any other bright source. A net
total count rate of $\sim$$1.4\times10^{-2}$ cts s$^{-1}$ after background
subtraction was measured for the whole energy range in {\em NUCLEUS}. Spectral
analysis was performed with {\em XSPEC} v.12.7 \citep{Arnaud96}, grouping source
counts to have at least 15 counts per bin in order to be able to apply
$\chi^{2}$ statistics in the fitting analysis. The final binned X-ray spectrum
of Haro~2 is shown in Fig.~\ref{figxspec}.

\begin{figure*}
\centering
\includegraphics[width=16cm,bb=36 248  573 544 dpi, clip=true]
{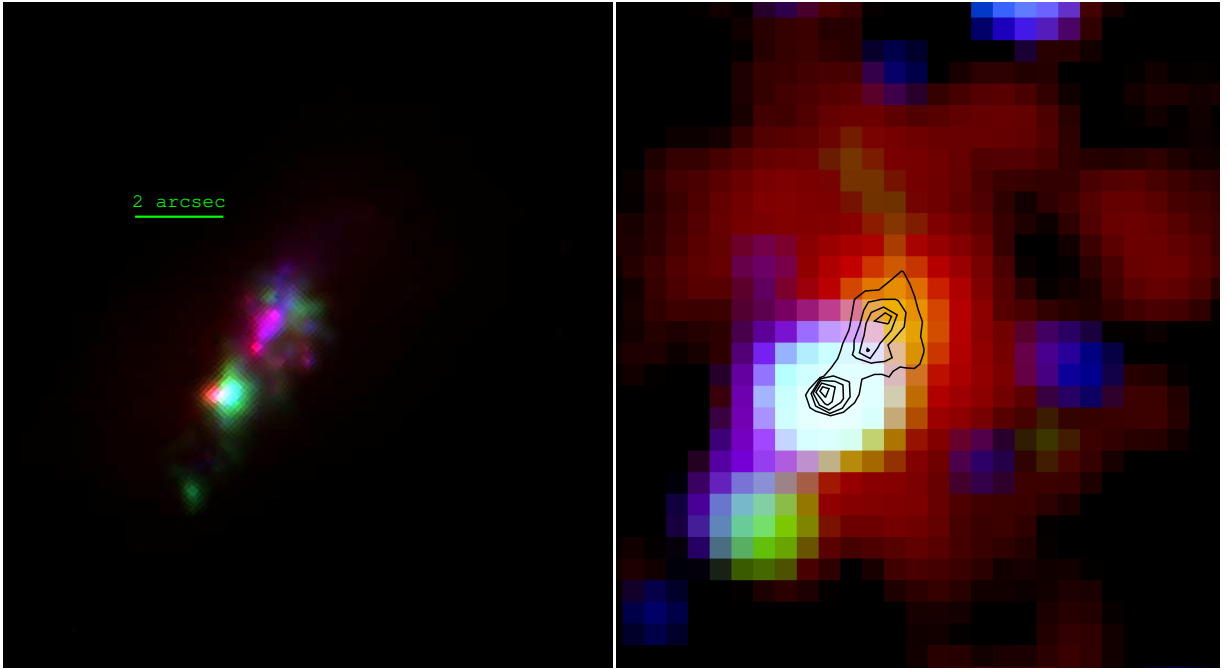}
\caption{Left: Multiwavelength image of Haro~2, as in Fig.~\ref{fignicwf}.  Right: the color-coded
{\em Chandra}/ACIS X-ray image (colors as in Fig.~\ref{figx}) at the same scale, with the NICMOS contours superimposed for
reference. North is up and East is left.}
\label{fignicwfx}
\end{figure*}

{\em Chandra}/ACIS X-ray image is displayed in Fig.~\ref{figx} colour-coded as a
function of the energy, where red: $0.2-1.5$ keV, green: $1.5-2.5$ keV and blue:
$2.5-8.0$ keV. Color-coded X-ray image is also shown in Fig.~\ref{fignicwfx}
with higher detail, together with the composite multiwavelength image at the
same scale for reference. We have also superimposed the NICMOS near-infrared
contours to identify the positions of the star-forming knots. Haro~2 shows a
diffuse soft X-ray emission extended mostly to the N and W of the starbursts,
over a region with radius $\sim$$600$ pc. Also, three hard, point-like sources are
observed in Fig.~\ref{figx} and labeled as X1, X2 and X3. Background-corrected
count rates of these hard sources for several energy bands are shown in
Table~\ref{xrayrate}, together with the values for {\em NUCLEUS}, which includes X1
and X2. X1 is located at the position of knot SE, and is associated to its
massive stellar cluster, as we will discuss later. X2 is a source $5$ times
weaker in the $1.5-8.0$ keV band than X1, located to the Southeast of knot SE. X2
does not have any correlation to any continuum or \hal\ emitting region, since,
as Fig.~\ref{fignicwfx} shows, the closer UV-optical bright knot is
$\sim$$1.5\arcsec$ far. An extra hard object X3 detached from the starburst region
is found $\sim$$10\arcsec$ Northwest from the nucleus, with a net flux of
$\sim$$1.5\times10^{-3}$ cts s$^{-1}$. The spectral image obtained with {\em
NOT}/ISIS in direction D2 (see Sect.~\ref{isis}) indicates that the ratio
\oxiil/\hb\ in the region close to X3 is two orders of magnitude higher than in
the nuclear region. This fact points towards X3 being a source within Haro~2
rather than a background source. However, given its angular distance from the
bulk of the X-ray emission and its low flux, we did not analyze it in this work.
Therefore, X1 is the dominant component of the hard emission in the {\em NUCLEUS}.

\begin{table*}
\caption{Count rates of X-ray objects corrected for background contamination.}
\label{xrayrate}
\centering
\begin{tabular}{lccc}
\hline\hline\\
Object & Total & Soft($0.2-1.5$ keV) & Medium($1.5-2.5$ keV) + Hard($2.5-8.0$ keV)\\
& (s$^{-1}$) & (s$^{-1}$) & (s$^{-1}$) \\
\hline\\
{\em NUCLEUS} & $2.0\times10^{-2}$ & $1.2\times10^{-2}$ & $3.2\times10^{-3}$ \\
X1 & $6.8\times10^{-3}$ & $4.2\times10^{-3}$ & $2.1\times10^{-3}$ \\
X2 & $8.9\times10^{-4}$ & $3.6\times10^{-4}$ & $4.2\times10^{-4}$ \\
X3 & $1.5\times10^{-3}$ & $4.2\times10^{-4}$ & $6.3\times10^{-4}$ \\
\hline
\end{tabular}
\end{table*}

\subsection{Ground-based observations} 

\label{isis}

An image of Haro~2 obtained with instrument ALFOSC on the {\em Nordic Optical
 Telescope} with filter V\_Bes 530\_80 is displayed in Fig.~\ref{figharo}
(courtesy of Luz Marina Cair\'os and Ricardo Amor\'in). Also, two long-slit
observations with ISIS on the {\em William Herschel Telescope} were downloaded
from the archive for which information is included in
Table~\ref{isistable}. ISIS spectral images of the source were acquired with
slit aligned along directions D1 (red and blue arm in December 25, $2006$) and
D2 (blue arm in February 9, $2002$). Although the observation along D2 was
performed only with the blue arm of the instrument, a similar spectral range
than the one achieved for direction D1 with both arms was obtained using an
observational mode with lower spectral dispersion. As a result, spatial
configuration of stellar continuum and main nebular emission lines was obtained
along both the major and minor axes of Haro~2. Specifically, we derived from
these images the spatial profiles of the \hb\ and \hal\ emission lines.

\section{Analysis and results}

\label{results}

\subsection{{\em HST} morphology and photometry}

The UV image of Haro~2 nucleus with the identification of its brightest
star-forming knots is shown in Fig.~\ref{figharo}, together with an image of the
whole galaxy in the $V$-band. The elliptically-shaped nucleus of Haro~2 hosts various massive
clusters, located on knots SE and NW. \citet{Mendez00} proposed to differentiate
2 smaller knots within NW, as indeed suggested by
Fig.~\ref{figuvopt-halpha}. Moreover, \citet{Chandar04} argued they could
identify 6 faint clusters within the galaxy, from the inspection of the UV
spatial profile along slit D2 (see Fig.~\ref{figstispro}).  However, for the
purposes of this work we will consider only knots NW and SE as individual
star-forming regions. These knots are aligned along the major axis, being
separated by $\sim$$75-100$ pc. While knot SE is not resolved, knot NW is more
extended and shows a more complex morphology. UV continuum extends few hundreds
of parsec away from the knots, so that the whole nuclear star-forming region has
a size of $\sim$$750\times 1000$ pc$^{2}$.

The UV-flux values for each knot and for the whole starburst region were extracted
from the FOC image considering boxes shown in Fig.~\ref{figuvopt-halpha} with sizes
$7.4\arcsec \times 5.5\arcsec$ for knot NW and $7.4\arcsec \times 4.8\arcsec$
for knot SE. As explained above, due to the proximity between knots NW and SE
the origin of the counts could not be discerned unambiguously, hence the total
flux of the whole source should be taken as a more reliable
measure. \fuv\ values for each knot together with their sizes are included in
Table~\ref{knotstable}. The total observed UV emission of the whole galaxy
nucleus detected by FOC is \fuv$=2.7\times10^{-14}$ \ergsacm\ which agrees with
the value from the spectrum taken by {\em IUE} with a total aperture of
$10\arcsec \times 20\arcsec$, which should enclose the whole UV-emitting
region. As it will be explained later, masses of the bursts were calculated
through the UV flux of each of the knots using evolutionary population
synthesis models.

Comparison of NICMOS image of IR continuum at $1.60$ $\mu$m with the WFPC-2
image of the UV continuum at $3300$ \AA\ (see Fig.~\ref{fignicwf}) shows that
some regions which are very prominent in the IR are not visible in the UV. This
results especially evident on knot SE, as well as in some more diffuse areas of
knot NW. Since both knots are dominated by young, massive stars, the
non-detection in the UV of the stars detected in the IR is probably due to a
patchy distribution of dust, which hides part of the young population, or to
evolutionary effects on different populations.

The \hal\ image obtained with {\em HST}/WFPC2 is shown in Fig.~\ref{figuvopt-halpha}
(right), with the UV continuum contours superimposed. The \hal\ emission appears
associated and essentially co-spatial to both stellar clusters SE and NW.  

\begin{table*}
\caption{Physical properties of knots NW and SE in Haro~2, assuming a
  single stellar population at each knot, with mass normalization
  according to their integrated UV luminosity (see text for details).}
\label{knotstable}
\centering
\begin{tabular}{lcccccc}
\hline\hline\\
Knot & \multicolumn{1}{c}{Age} & \multicolumn{1}{c}{Mass} & \ebv$^{\mathrm{a}}$ &
\multicolumn{1}{c}{\fuv} & \multicolumn{1}{c}{\luv$^{\mathrm{b}}$} &
\multicolumn{1}{c}{Stellar extension$^{\mathrm{c}}$} \\
& (Myr) & (\msun) & & (\ergsacm) & (\ergsa) & (pc) \\
\hline\\
Knot SE & $\sim$$4$ & $7.0\times 10^{5}$ & $0.050$ & $1.1\times 10^{-14}$ & $5.5\times
10^{38}$ & $\sim$$60$ \\
Knot NW & $\sim$$5$ & $1.3\times 10^{6}$ & $0.020$ & $1.7\times 10^{-14}$ & $8.5\times
10^{38}$ & $\sim$$150$ \\  
\hline
\end{tabular}
\begin{list}{}{}
\item[$^{\mathrm{a}}$] Internal color excess assuming \citet{Prevot84} extinction law
(SMC). 
\item[$^{\mathrm{b}}$] Luminosity value not corrected for Galactic or intrinsic extinctions.
\item[$^{\mathrm{c}}$] Extension of the stellar continuum along direction D2 in the STIS
spectral image.
\end{list}
\end{table*}

\begin{table*}
\caption{Predicted values of the ages, masses, stellar and nebular extinctions and emissions of Populations 1, 2 and the composite of both, as well as the \ewhb\ produced. Observational values are also displayed.}
\label{poptable}
\centering
\begin{tabular}{ccccccccc}
\hline\hline\\
Population & $Age$ & $M$ & \ebv$_{*}$ & \ebv$_{neb}$ & \luv$^{\mathrm{a}}$ & \lha$^{\mathrm{a}}$ & \lfir\ & \ewhb\  \\
 & (Myr) & (\msun) &  &  & (\ergsa) & (\ergs) & (\ergs) & (\AA) \\
\hline\\
1 & $4.5$ & $1.8\times 10^{6}$ & $0.035$ & $0.24$ & $1.3\times 10^{39}$ & $2.5\times 10^{40}$ & $3.4\times 10^{42}$ & - \\ 
2 & $3.5$ & $2.0\times 10^{6}$ & $0.5$ & $0.24$ & $7.1\times 10^{37}$ & $5.3\times 10^{40}$ & $1.6\times 10^{43}$ & - \\
1+2 & - & $3.8\times 10^{6}$ & - & $0.24$ & $1.4\times 10^{39}$ & $7.8\times 10^{40}$ & $2.0\times 10^{43}$ & $58$ \\
\hline\\
Observed & - & - & - & $0.24^{\mathrm{b}}$ & $1.4\times 10^{39}$ & $7.8\times 10^{40}$ $^{\mathrm{b}}$ & $2.0\times 10^{43}$ & $44-80^{\mathrm{b}}$ \\  
\hline
\end{tabular}
\begin{list}{}{}
\item[$^{\mathrm{a}}$] Luminosity value reddened by Galactic and intrinsic extinctions.
\item[$^{\mathrm{b}}$] Value from \citet{Moustakas06}.
\end{list}
\end{table*}

\subsection{UV continuum}

\label{resuvcont}

\begin{figure*}
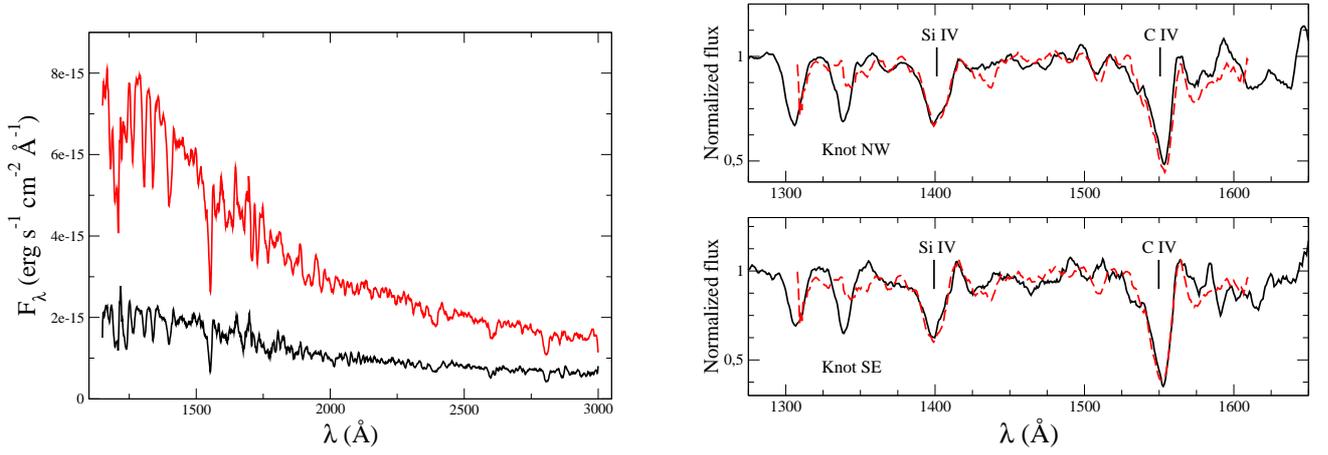

\begin{center}
\includegraphics[width=8cm,bb=2 34 707 522 dpi,clip=true]{scaled_spectra.k1.k2.07.eps}
\hspace{1.0 cm}
\includegraphics[width=8cm,bb=42 34 702 522
dpi,clip=true]{normalized_spectrum.knot1_knot2.stracksZ008_uvZ020.02.eps}
\end{center}
\caption{Left: observed UV stellar continuum of knot SE ({\em bottom}) and knot NW ({\em
top}). Right: normalized spectrum of UV stellar continuum of knots SE and NW (solid lines)
with absorption lines \siivd\ and \civd\ fitted with SB99 models for ages $\sim$$4$ Myr and
$\sim$$5$ Myr (dashed lines), respectively (see text for details).}
\label{figcontuv}
\end{figure*}

High spectral resolution UV spectra are available in Starburst99 models
(\citet{Leitherer99}, hereinafter SB99 models) in the range $1200-1600$ \AA\ for the
LMC/SMC library and $1200-1800$ \AA\ for the Milky Way library, with a spectral step of
$0.75$ \AA. Normalized spectra for both knots in the wavelength range $1150-1700$ \AA\
were extracted from STIS spectral image obtained in the direction D2 with filter G140L. The
spatial extension of the stellar continuum regions considered was $0.6\arcsec$ and
$1.5\arcsec$ for knots SE and NW, respectively. Given the sentivity of the shape and
intensity of the absorption lines \siivd\ and \civd\ to the age of the burst, normalized
UV spectra from both knots were fitted individually with SB99 UV models in order to
calculate their evolutionary state. SB99 models were computed for an initial mass function
$\phi(m)$$\sim$$m^{-2.35}$ (i.e. a Salpeter IMF) and mass limits of $2-120$ \msun,
assuming instantaneous star formation and $Z=0.008$ for the evolutionary tracks and
high-resolution spectra. This metallicity is the closest available to the observational
one $Z$$\sim$$Z_\odot/2$ \citep{Davidge89}. UV line spectra from the models were either
taken from the Milky Way or the LMC/SMC libraries. We found that no acceptable fitting of
the normalized spectrum of either knot was possible with the LMC/SMC library whereas a
good agreement was reached when the Milky Way library was used. We obtained ages of
$\sim$$4$ Myr and $\sim$$5$ Myr for knots SE and NW. Uncertainties in the procedure may
yield a precision in the values of the ages obtained of approximately $\pm0.5$ Myr.  The
final fittings of the normalized spectra are shown in Fig.~\ref{figcontuv}.

Reddening in Haro~2 was calculated through the fitting of the UV SED of each
knot with the SB99 models, given the assumptions for IMF and metallicity already
explained above. SEDs of the models at the ages $4$ Myr (for knot SE) and $5$
Myr (for knot NW) were reddened for a certain value of the color excess \ebv\
and assuming SMC extinction law \citep{Prevot84} and $R_{V}=3.1$, then they were
redshifted and finally extinguished for Galactic extinction for the color excess
\ebv$=0.012$ assuming Cardelli extinction law \citep{Cardelli89}. Redshift was
obtained from the Gaussian fit of the \hb\ line included in the spectrum of the
G430L observation, which turned out to be $z=0.00461$, that is within
$\sim$$4$\% the value reported in NED (NASA/IPAC Extragalactic Database).
Iterative method for calculating \ebv\ was repeated until a good fit was
achieved for each knot, obtaining \ebv$_{\rm{knot\,SE}}$$\sim$$0.050$ and
\ebv$_{\rm{knot\,NW}}$$\sim$$0.020$. Observed spectra are shown in
Fig.~\ref{figcontuv}, together with the fitting SEDs of the models reddened.
Finally, the masses of both knots were calculated dividing the observed UV
luminosity \luv\ of each knot measured by {\em HST}/FOC by the reddened,
redshifted, predicted value from the SB99 models, which is computed per unit
mass, obtaining $M_{\rm{knot\,SE}}$$\sim$$7\times10^{5}$ \msun\ and
$M_{\rm{knot\,NW}}$$\sim$$1.3\times10^{6}$ \msun. These values are obtained when
considering IMF mass limits of $2-120$ \msun. Were other boundaries of the mass
range be considered, a proper correction should be performed to these values
\citep{Wilkins08,Oti10}. Given the difficulties to discern the actual flux of
each knot from the total one, the values of the masses calculated should be
taken as a rough estimate. As explained in Sect.2.1.1, this procedure
assumes that the stars contributing to the diffuse UV continuum within each box,
as indicated in Fig.~\ref{figuvopt-halpha}, share the same age as those in the
central knots dominating the UV spectral features. While this should be
considered just as a first order approximation, this strategy provides a more
realistic global mass normalization than just considering the flux within the
unresolved knots. A correct normalization is required to properly compare the predictions with globally
integrated magnitudes, such as the X-ray or far infrared luminosities.

Physical properties calculated through evolutionary population synthesis models
for knots SE and NW are shown in Table \ref{knotstable}. Given the small volume
of knot SE, which we can not resolve, and its mass,
$M_{\rm{knot\,SE}}>1\times10^{5}$ \msun, this knot should be considered a super
stellar cluster \citep{Adamo10}.

We want to stress that the results of the fit discussed above are based on the
UV continuum and spectral features as observed with STIS. \lfir\ and \lha\
emissions provide us with a consistency check to verify the validity of the
fitting. Once the parameters of the models were calculated, we compared their
predictions of \lfir\ and \lha\ with the observational values. Observed \lfir\
luminosity was calculated through the relation of \citet{Helou85} given the {\em
IRAS} fluxes $F_{60\; \mu \rm{m}}=4.7$ Jy and $F_{100\; \mu \rm{m}}=5.3$ Jy.
Since this relation yields the infrared emission collected in the wavelength
range $42.5-122.5$ $\mu$m, it should be corrected in order to obtain the total
emission in the complete infrared range. As was found by \citet{Calzetti00} for
a sample of eight starburst galaxies, the ratio $FIR(1-1000\; \mu
\rm{m})/FIR(40-120\; \mu \rm{m})$$\sim$$1.75$, hence the value obtained from
\citet{Helou85} relation was multiplied by $1.75$, yielding
\lfir$=2.0\times10^{43}$ \ergs. On the other hand, \citet{Moustakas06} measured
the integrated intensities and the equivalent widths of the Balmer and forbidden
lines for more than 400 galaxies using drizzled spectroscopic observations with
Boller \& Chivens spectrograph on Bok telescope. Intensities and equivalent
widths of all lines were corrected for Galactic reddening and stellar
absorption, and also for $[NII]$ contamination in the case of \hal, obtaining
for Haro~2 $F$(\hal)$=1.6\times10^{-12}$ \ergscm. Since their extraction
aperture was of $50\arcsec$ and the slit was drifted along $40\arcsec$, this
value is expected to represent well the total flux associated to the burst. The
same $F$(\hal) value for Haro~2 was measured by \citet{GilDePaz03}. Moreover,
besides the integrated values, \citet{Moustakas06} reported the values of
nuclear regions using an aperture of $2.5\arcsec \times 2.5\arcsec$. Nebular
reddening calculated through the ratios of the integrated Balmer lines values
from \citet{Moustakas06} is \ebv=$0.24$, which is in perfect agreement with the
value \ebv=$0.22$ reported by \citet{MasHesse99b}. This value is $\sim$$0.2$ dex
larger than the values found by continuum SED fitting for knots SE and NW.
Higher extinction for nebular gas than for stellar continuum has been observed
in several star-forming regions (see \citet{Maiz1998} and \citet{Calzetti00}).
This disagreement is usually attributed to the fact that young clusters clean
its surroundings from dust, leaving relatively clear holes through which the UV
continuum escapes with low attenuation. On the other hand, the more extended
ionized gas still contains a significant amount of dust.

\lfir\ and \lha\ were calculated with the evolutionary synthesis models by
\citet{Cervino02b} (hereinafter CMHK02 models). \lfir\ is produced by the emission of
dust, which is assumed to be in thermal equilibrium.  The models compute the energy
absorbed by dust, comparing the intrinsic UV-optical continuum with the observed one, and
assumes that all the missing energy is reemitted as \lfir. In addition, the models assume
that  a fraction $1-f$$=$$0.3$ of the ionizing photons is directly absorbed by dust and
does not contribute to the ionization. \lfir\ and \lha\ were estimated with the CMHK02
models for the knots SE and NW, assuming the values (mass, age, reddening, ...) obtained
in the previous analysis and shown in Table~\ref{knotstable}. When the values of \lfir\
and \lha\ were compared to the observational ones we found that the models severely
underestimated both the \lfir\ and \lha\ luminosities by factors $7$ and $3$,
respectively.

This disagreement points to the existence of an extremely reddened stellar
population, whose contribution to the UV continuum is negligible in the FOC
image and the STIS spectra, but which contributes significantly to both the
observed FIR and \hal\ emissions. Moreover, the high underestimation of \lha\
points to a young population, with a relatively strong ionizing power.
Therefore, in order to explain all observables we built a simplified model based
on the existence of two populations: Population 1, which appears as a
conspicuous UV-emitter, and Population 2, which is highly reddened and is the
main contributor to \lfir\ and \lha. We defined three equations with the
observables UV, FIR and \hal\ fluxes being the sum of the reddened contributions
by each of the populations, where their masses $M_{1}$ and $M_{2}$ and the
stellar extinction of Population 2 \ebv$_{*,2}$ are the unknowns. The
contribution by each population was calculated with the CMHK02 models assuming
the corresponding parameter values. A mean age of $Age(1)=4.5$ Myr and stellar
reddening \ebv$_{*,1}=0.035$ were assumed for Population 1, combining the
results for both knots NW and SE and mean nebular extinction was assumed to be
\ebv$_{neb}$=$0.24$ \citep{MasHesse99b,Moustakas06} irrespective of the
population. For each age of Population 2 considered ($Age(2)$), a system of
three equations with three unkowns was obtained and solved. The results show
that in order to reproduce the observed values, stars from Population 2 must be
in a less evolved state than those of Population 1, which agrees with the former
being more reddened having had less time to clean the surroundings. Values
$Age(2)=3.5$ Myr and stellar color excess \ebv$_{*,2}=0.5$ can account for the
total UV, \hal\ and \lfir\ emissions. Assuming lower values for \ebv$_{*,2}$
implies a higher contribution by Population 2 to the UV emission, which as we
have discussed is ruled out by the FOC image and the
STIS spectrum. Values found for $M_{1}$, $M_{2}$ and \ebv$_{*,1-2}$ are included
in Table~\ref{poptable} together with the ages considered, the contribution
by each population to the magnitudes observed and the global \ewhb. Values from
Table~\ref{knotstable} do differ from those of Population 1 of
Table~\ref{poptable} since different assumptions were made in each case. The
values given in Table~\ref{knotstable} correspond to the average properties of
the subsample of massive stars which are less affected by interestellar
extinction, while a similar fraction of young stars seems to be still too obscured
to be detected in the UV. While this is a simplified scenario, it shows that in
reality the knots of Haro~2 are probably formed by young massive stars with an
age within $\sim$$3.5$--$5$ Myr, and a continuous sequence of interstellar
extinction, spanning from regions almost devoid of dust to others mostly blocked
in the UV by dense dust clouds.

The underestimation of the Balmer and FIR emissions in Haro~2 when assuming a starburst
mass derived from the UV flux was also found by \citet{MasHesse99b} who,
assuming a single reddened stellar population, obtained lower \lhb\ and
\lfir\ values than observed by factors $2.3$ and $2$. These factors differ from
ours since \citet{MasHesse99b} considered a steeper IMF with slope
$\alpha=-3$. On the contrary, they obtained an age for the whole nuclear
star-forming region of $4.8$ Myr using $EW($\hb$)$ corrected for both reddening
and underlying stellar absorption, which agrees with our estimated value. Using
an updated set of models \citet{Cervino02b} found a value of $4.2$ Myr assuming
the very same value of $EW($\hb$)$ from \citet{MasHesse99b}. \citet{Chandar04}
fitted the UV spectrum from G140L observation along direction D2 using SB99
models and giving different weights to continuum, interstellar lines and stellar
wind lines, obtaining the best fit for an age of $5\pm1$ Myr. In all these cases
the study was performed over the whole galaxy. Moreover, \citet{Moustakas06}
obtained $EW($\hb$)=44$ \AA\ and $EW($\hb$)=80$ \AA\ for the whole galaxy and
the nuclear region, respectively. Both values are already corrected for the
Galactic and internal extinctions and stellar absorption. Our prediction from
the models when assuming the existence of Populations~1 and 2 of $EW($\hb$)$$\sim$$58$ \AA\ lies between both values.

\subsection{\lyalp\ emission}

\label{lyaemission}

\begin{figure*}
\begin{center}
\includegraphics[width=13cm,bb=0 104 720 540 dpi,clip=true]{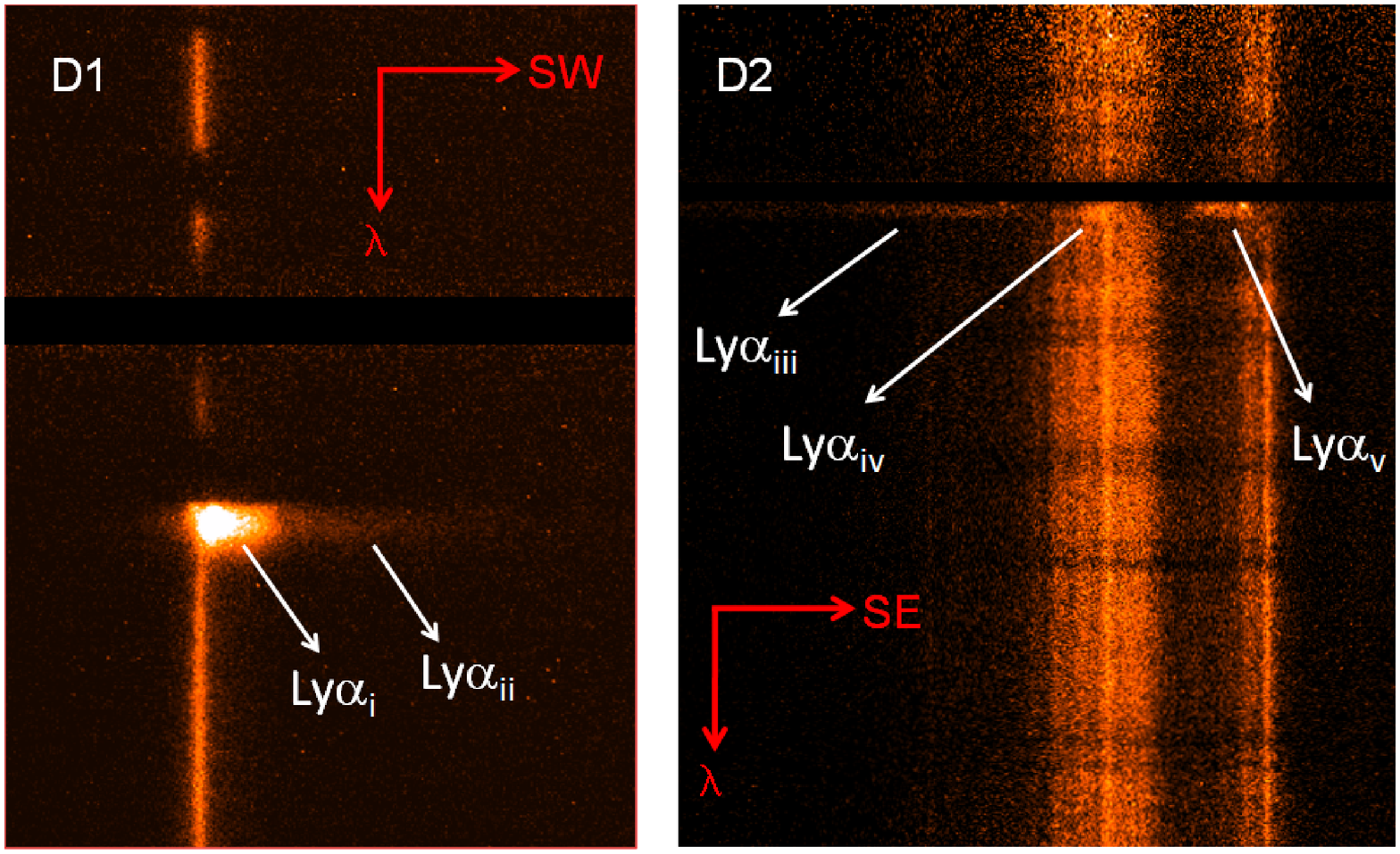}
\includegraphics[width=16cm,bb=13 47 785 528
dpi,clip=true]{ratios_lyalpha_halpha_hbeta.leitherer_kunth.15.eps}
\end{center}
\caption{{\bf Left}: slit along direction D1. {\bf \em Top-left}: {\em HST}/STIS G140M
spectral image of Haro~2. {\bf \em Middle-left}: Emission profiles of \lyalp\ (blue) and
UV continuum ($1175$ \AA\ rest frame, black) from STIS data, and \hal\ (red) from WFPC2
data. {\bf \em Bottom-left}: Spatial profiles of \lyalp\ (blue) and observed
\lyalp$/$\hal\ (black) and \hal$/$\hb\ (from {\em WHT}/ISIS data, in red). {\bf Right}:
slit along direction D2. {\bf \em Top-right}: {\em HST}/STIS G140L spectral image. {\bf
\em Middle-right}: Emission profiles of \lyalp\ (blue) and UV continuum ($1227$ \AA,
black) from STIS data, and \hal\ (red) from WFPC2 data. {\bf \em Bottom-right}: Spatial
profiles of \lyalp\ (blue) and observed \lyalp$/$\hal\ (black) and \hal$/$\hb\ (from {\em
WHT}/ISIS data, in red). The geocoronal \lyalp\ line has been blackened in both spectral
images at the top. In the middle and bottom panels positive $x$-axis corresponds to
Southwest (left) and Southeast (right). Vertical scale in the middle panels corresponds to
the \lyalp\ and \hal\ fluxes. To ease the comparison, the dashed line shows the \hal\
profile scaled by a factor 0.4 (left) and 0.085 (right). The vertical scale in the
bottom panels corresponds to the \lyalp$/$\hal\ and \hal$/$\hb\ ratios. The brown dashed
lines represent the expected \lyalp$/$\hal\ ratio, assuming case B recombination and
applying the internal reddening derived from the observed \hal$/$\hb\ ratio (Balmer
decrement), as well as the Galactic extinction.}
\label{figstispro}
\end{figure*}

Spectral images on the ultraviolet along the minor (D1 direction) and major (D2 direction)
axes in the \lyalp\ wavelength range are displayed in Fig.~\ref{figstispro} and a
relatively strong \lyalp\ emission is observed in both cases, extending well beyond the
stellar clusters. In the case of the observation of knot SE along D1 direction the
\lyalp-emitting region extends largely to the southwest direction whereas it is much
weaker Northeast of the stellar cluster. \citet{MasHesse03} showed that the spatial
extension of this \lyalp\ emission along this direction is much larger than the
continuum-emitting region. Even more, as observed in Fig.~\ref{figstispro} there seems to
be a strong, compact component near the stellar continuum, which extends over $\sim$$250$
pc (\lyai, hereinafter, $F($\lyalp$)=2.0\times10^{-13}$ \ergscm\ not corrected for any
kind of absorption), and another weaker and more diffuse emission over $\sim$$850$ pc
extending to the southwest direction (\lyaii, $F($\lyalp$)=2.8\times10^{-14}$ \ergscm).

In the spectral image along D2 direction displayed in Fig.~\ref{figstispro}
\lyalp\ emission seems to be more complex, with regions of total absorption and
total emission.  Fig.~\ref{figstispro} shows the spatial profile along direction
D2 of \lyalp, being its value positive whenever it appears as emission and
negative were it pure absorption. The UV stellar continuum from {\em HST}/STIS
is superimposed to the \lyalp-profile and a clear spatial detachment is observed
between both emissions. Whereas the stellar continuum shows two peaks, one
rather narrow due to the knot SE ($\sim$$60$ pc) and the other one more extended
owing to knot NW ($\sim$$150$ pc), \lyalp-profile is found in emission in three
different regions: 1) Northwest of knot NW, totally detached from the stellar
continuum and extending over $\geq600$ pc (emission \lyaiii,
$F($\lyalp$)=1.8\times10^{-14}$ \ergscm), 2) within knot NW, $<50$ pc (emission
\lyaiv, $F($\lyalp$)=5\times10^{-15}$ \ergscm), and 3) Northwest from knot SE
and $\sim$$100$ pc long (emission \lyav, $F($\lyalp$)=1.6\times10^{-14}$
\ergscm). While \lyaiii\ shows a diffuse morphology, the \lyaiv\ and \lyav\
contributions seem rather compact. However, the diffuse emission \lyaiii\,
although dimmer than the other two components, is the strongest one when
integrated over its large extension, since it is observed up to the northwestern
extreme of the slit. \citet{Hayes07} found also in Haro~11 that the diffuse
emission surrounding the stellar clusters is the most prominent contribution to
\lulyalp.  On the other hand, emission \lyaiv\ is the weakest component, and
unlike emissions \lyav\ and \lyaiii\ it is fully located within one stellar
knot.

\subsection{X-rays}

\label{resxrays}

As explained in Sect.~\ref{obsxrays} the spectrum of the region labeled as {\em NUCLEUS} in Fig.~\ref{figx} was analyzed. This region includes hard sources X1 and X2, with the former completely dominating the hard emission of {\em NUCLEUS}. Fitting of the spectrum was performed with XSPEC
going from simpler to more complex models, evaluating the statistical
significance of the change at each step through the F-test and considering it
valid if probability significance was $>99$\%. Galactic absorption was fixed to
$N($HI$)_{Gal}=6.3\times 10^{19}$ cm$^{-2}$, whereas the intrinsic absorption
was fixed to $N($HI$)_{int}=7.0\times 10^{19}$ cm$^{-2}$ as derived by
\citet{Lequeux95}. When freeing the hydrogen column density of the intrinsic
absorption it was found that the result of the fit and the value of
$N($HI$)_{int}$ did not change significantly, therefore we decided to fix
it. For the metallicity we assumed the value $Z=Z_\odot/2$ obtained by
\citet{Davidge89}. The model which yielded a better fit is a composite of a hot
plasma gas (hereinafter, HP, corresponding to the {\tt mekal} model in XSPEC \citep{Mewe85}) and a power-law emission (PL, in what follows)
attenuated by both intrinsic and Galactic absorptions. HP would correspond to
the diffuse soft emission filling the starbursting region and its surroundings,
while the PL component is mostly dominated by X1. For the best fit achieved the
temperature of the HP is $kT=0.71_{-0.10}^{+0.11}$ keV and the PL index
$\Gamma=1.8_{-0.4}^{+0.3}$. No acceptable fit is obtained when considering only one single emitting component. Results assuming a Raymond-Smith model ({\tt raymond} model in XSPEC \citep{Raymond77}) for the HP instead of {\tt mekal} yielded similar results.

\begin{table*}
\caption{X-ray spectral fitting.}
\label{xrayfit}
\centering
\begin{tabular}{lccccc}
\hline\hline\\
Model & \multicolumn{1}{c}{$kT$ (keV)} & $\Gamma$ & Metallicity & $\nu$ & $\chi^{2}/\nu$ 
 \vspace*{1 mm} \\
& Norm$^{\mathrm{a}}$ & Norm$^{\mathrm{a}}$ & $Z_\odot$ & & \\
\hline\\
1: wabs*zwabs*mekal & \multicolumn{1}{c}{$0.81$} & \multicolumn{1}{c}{-} & $0.5$ & $16$ & $3.486$ \\
& $2.9\times10^{-5}$ & - & & & \\
2: wabs*zwabs*zpowerlw & - & \multicolumn{1}{c}{$2.1$} & - & $16$ & $2.947$ \\
& - & $1.6\times10^{-5}$ & & & \\
3: wabs*zwabs*(mekal+zpowerlw) & \multicolumn{1}{c}{$0.71_{-0.10}^{+0.11}$} & \multicolumn{1}{c}{$1.8_{-0.4}^{+0.3}$} & $0.5$ & $14$ &
$1.083$ \vspace*{1 mm} \\ 
& $1.6_{-0.5}^{+0.5}\times10^{-5}$ & $9_{-3}^{+3}\times10^{-6}$ & & & \vspace*{1 mm} \\
\hline
\end{tabular}
\begin{list}{}{}
\item[$^{\mathrm{a}}$] Units of normalization as in XSPEC. Hot plasma: $10^{-14}/\{4\pi [D_{A} (1+z)]^{2}\} \int n_{e} n_{H} \rm{d} V$. Power-law: photons s$^{-1}$ cm$^{-2}$ keV$^{-1}$ ($1$ keV).
\end{list}
\end{table*}

\begin{table*}
\caption{X-ray fluxes and luminosities.}
\label{xraylum}
\centering
\begin{tabular}{lccccc}
\hline\hline\\
Model & $F_{0.4-2.4\rm{\,keV}}$ $^{\mathrm{a}}$ & $F_{2.0-10.0\rm{\,keV}}$ $^{\mathrm{a}}$ & $L_{0.4-2.4\rm{\,keV}}$ $^{\mathrm{b}}$ & $L_{2.0-10.0\rm{\,keV}}$ $^{\mathrm{b}}$ & ($L_{0.4-2.4\rm{\,keV}}^{PL}/L_{0.4-2.4\rm{\,keV}}^{HP}$) $^{\mathrm{b}}$ \vspace*{1 mm} \\
& (\ergscm) & (\ergscm) & (\ergs) & (\ergs) & \\
\hline\\
3: wabs*zwabs*(mekal+zpowerlw) & $4.6_{-1.1}^{+1.0}\times10^{-14}$ & $3.1_{-1.5}^{+1.7}\times10^{-14}$ & $2.4_{-0.6}^{+0.5}\times10^{39}$ & $1.5_{-0.8}^{+0.8}\times10^{39}$ & $1.2_{-0.5}^{+0.4}$ \vspace*{1 mm} \\ 
\hline
\end{tabular}
\begin{list}{}{}
\item[$^{\mathrm{a}}$] Values of fluxes have not been corrected for neutral absorption.
\item[$^{\mathrm{b}}$] Values of luminosities have been corrected for neutral absorption.
\end{list}
\end{table*}

Other options were tested, like fitting the spectrum with a
two-temperature hot plasma, but large uncertainties were obtained for the higher
temperature. Also, metal abundances were freed after modeling the HP with {\tt
vmekal} \citep{Mewe85}, obtaining best fits when O and Mg abundances were $Z(O)/Z(O)_\odot$$\sim$$2.5$
and $Z(Mg)/Z(Mg)_\odot$$\sim$$4$. However these models led to values of the reduced
chi-squared of $\chi^{2}/\nu$$\sim$$0.7$, which indicates an overparameterization of
the fitting and were therefore rejected. On the other hand, as explained by
\citet{Ott05A}, $\alpha$-elements are injected into the gas surrounding the
stellar clusters by Type II SNe, whereas iron is mainly yielded into the
interstellar medium (ISM) by Type I SNe, causing a time delay between both
enrichments due to the different evolution of both types of stars producing each
class of supernova. This causes a supersolar value of the ratio
$Z(\alpha)/Z(\rm{Fe})$ in the gas inside bubbles blown by starbursts which have
been enriched by supernovae yields. This effect was modeled by \citet{Silich01},
who predicted an oversolar abundance of O compared to Fe during more than $40$
Myr after the onset of the starburst. In order to study the overabundance of
$\alpha$-elements inside the hot gas we followed the procedure by
\citet{Martin02}: the HP emission was modeled by {\tt vmekal}, He was fixed to
the nebular abundance, Mg, Ne, Si and Ca were linked to the O abundance which was freed (considered to be the $\alpha$-group, $Z(\alpha)$), and
the rest of the metals abundances were fixed to Fe abundance which was also left
free (Fe group, $Z(Fe)$). However, no statistically significant improvement in
the fit was obtained. We studied the inclusion of S and Ar into the
$\alpha$-group as \citet{Grimes05} and \citet{Kobulnicky10} did, but very
similar results were obtained.

\begin{figure}
\centering
\includegraphics[width=8cm,bb=14 34 701 528
dpi,clip=true]{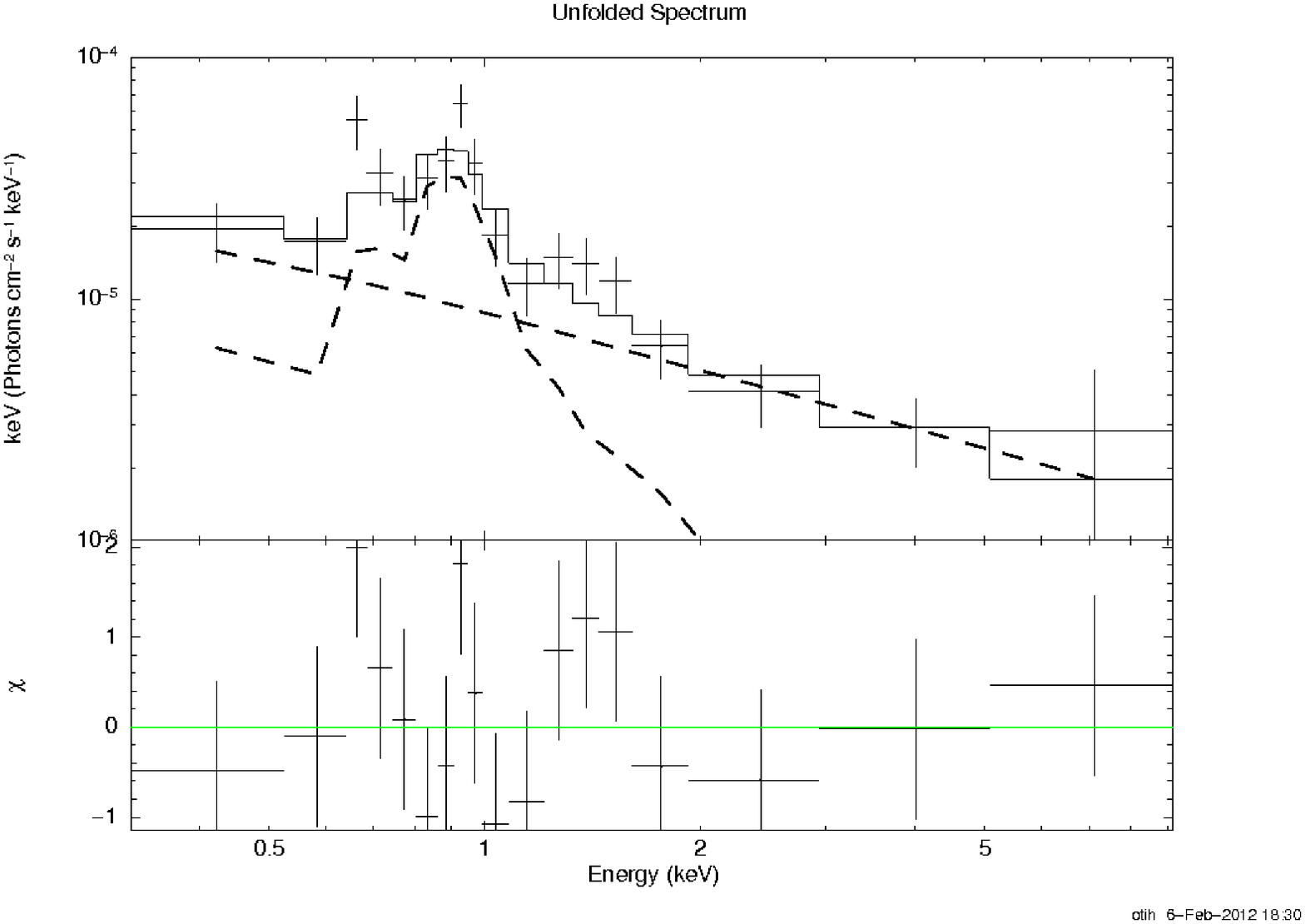}
\caption{X-ray spectrum of Haro~2 with the best fitting model (Model 3 in Table.~\ref{xrayfit}). Thermal and power-law components are shown in dashed lines. Fitting residuals are also shown in the bottom panel.}
\label{figxspec}
\end{figure}

The final results of the X-ray spectral fitting are shown in
Table~\ref{xrayfit}. Table~\ref{xraylum} contains the X-ray luminosities for the best fitting model (Model 3), as well as the ratio of the contribution of the PL component over the HP one in the soft X-ray luminosity $L_{0.4-2.4\rm{\,keV}}$. Finally Fig.~\ref{figxspec} shows the X-ray spectrum of Haro~2 together with Model 3. Table~\ref{xrayfit} lists the values of
the temperature $kT$ of the HP, the PL index $\Gamma$, the metallicity of the gas
assumed, the number of degrees of freedom $\nu$ and the reduced chi-squared
$\chi^{2}/\nu$. Errors correspond to
a confidence level of $90$\%. \citet{Stevens98} obtained a lower
temperature value $kT=0.36$ keV when analyzing the {\em ROSAT}/PSPC observation
of Haro~2. They assumed a very low metallicity $Z$$\sim$$Z_\odot/10$ and obtained a higher
intrinsic hydrogen column density of $N($HI$)=2.2\times10^{21}$ cm$^{-2}$ with very high uncertainties, which might
explain the disagreement. On the other hand the value we obtained in our study
for $kT$ is similar to the values found in other star-forming galaxies. For
instance, \citet{Summers04} obtained a value of $kT=0.75$ keV for the hotter
component in their two-temperature composite hot plasma when fitting the soft
X-ray emission of NGC 5253, and \citet{Martin02} obtained $kT=0.7$ keV for the
component of the disk in the dwarf starbursting galaxy NGC 1569. \citet{Hayes07}
and \citet{Grimes07} obtained $kT=0.69$ keV and $kT=0.68$ keV for Haro~11,
respectively, and \citet{Grimes05} and \citet{Kobulnicky10} calculated $kT=0.60$
keV and $kT=0.63$ keV for He 2-10, respectively. However, these values are
relatively high when compared to the study by \citet{Strickland04}, who obtained
$kT=0.11$ keV and $kT=0.37$ keV for a two-temperature model fitted simultaneously to
eight starbursts whose X-ray point-sources had been removed. Also,
\citet{Grimes05} only found two sources with $kT>0.5$ keV among seven dwarf
starbursts. The $kT$ value we propose for Haro~2 agrees better with the values
found by \citet{Grimes05} for their sample of ULIRGs.\\

As previously said, hard emission in Haro~2 is dominated by the point-like
source X1. Attending to \citet{Persic02} main contributors to the hard X-ray
energy range $2-15$ keV in starburst galaxies are High-Mass X-ray Binaries
(HMXB) accreting gas onto a compact object. The great spatial resolution of {\em
Chandra} enabled \citet{Strickland07} to extract the composite spectrum of the
detected X-ray point sources in the nuclear starbursting region of M82
(excluding M82 X1 identified as an UltraLuminous X-ray Source). After fitting
the hard energy range using a power law they obtained $\Gamma$$\sim$$1.0$. Similar
values were found by \citet{ShawGreening09} for 18 objects in M31
($\Gamma=0.8\mbox-1.2$). However, the fitting of Haro~2 X-ray spectrum
points towards a photon index value of $\Gamma$$\sim$$1.8$ for X1. Even more, the X-ray
luminosity of X1 ($L_{0.4-10.0\rm{\,keV}}=2.7\times10^{39}$ \ergs) is too large for a HMXB, and rather
classifies it as an UltraLuminous X-ray Source (ULX), a type of
hard-X-ray-emitting object with $L_{2.0-10.0\rm{\,keV}}\geq 10^{39}$ \ergs\ and
$\Gamma$ values in the range $1.6$ - $2.2$ \citep{Colbert04}. Several of these
sources have been highly studied in nearby galaxies where they can be spatially
separated from the surrounding components by {\em Chandra}. \citet{Kaaret01} and
\citet{Matsumoto01} reported the observation of the ULX M82 X-1, a compact, hard
X-ray emitter within the local starbursting galaxy M82 and $170$ pc far from its
dynamical center. They found that its total X-ray luminosity is $L_{X}>10^{40}$
\ergs\ and \citet{Matsumoto01} calculated that the probability of this source
being a background AGN is $<1$\%. When fitted with an absorbed power law
\citet{Kaaret06} obtained a photon index $\Gamma=1.67$ for this ULX.
\citet{Kong07} analyzed another hard, compact source in M82 labelled as X42.3+59
only $\sim$$5\arcsec$ ($\sim$$90$ pc) away from M82 X-1. They also identified it as an ULX and they
reported $\Gamma=1.3-1.7$ and $L_{2.0-10.0\rm{\,keV}}=(7.8-11)\times10^{39}$
\ergs. Even more, \citet{Strickland01} analyzed the X-ray emission of the
starburst galaxy NGC 3628, finding a compact object $\sim$$20\arcsec$ ($\sim$$1$ kpc) away from the
nucleus. They obtained that the model which best fitted the X-ray emission of the
object was an absorbed power law, finding $\Gamma=1.8$ and
$L_{0.3-8.0\rm{\,keV}}=1.1\times10^{40}$ \ergs. \citet{Grimes07} also suggested that the
hard, intense X-ray emitter they found analyzing a {\em Chandra} image of
Haro~11 (also a local \lyalp-emitting starburst galaxy as Haro~2), could be an ULX, with a
photon index $\Gamma=1.6$ and a luminosity of
$L_{0.3-8.0\rm{\,keV}}=2.2\times10^{39}$ \ergs. Indeed, the X-ray luminosity $L_{0.4-10.0\rm{\,keV}}=2.7\times10^{39}$ \ergs and
the photon index $\Gamma$$\sim$$1.8$ provided by the X-ray spectral fit of Haro~2 point
towards X1 being an ULX, rather than an ordinary HMXB.

\citet{Strickland07} found another source of hard X-rays in the star-forming
nucleus of M82. After removing the point sources of the region they identified a
diffuse hard X-ray component showing spectral features clearly different to the
composite spectrum of the point sources removed and which extends along the
plane of M82. They identified this component as the wind fluid driving the
superwind of the galaxy, emitting as much as
$L_{2.0-8.0\rm{\,keV}}=4.4\times10^{39}$ \ergs, and produced by the collimation
of the massive stellar winds and SN ejecta of the newly formed masive stars.
This emission can be spectrally fitted in the hard X-rays range either with a
Bremsstrahlung thermal model ($kT$$\sim$$3-4$ keV), or with a power law with
$\Gamma>2.0$ yielding similar results. Nevertheless, we conclude that a wind
fluid can not be responsible for the observed hard X-ray emission in Haro~2
since the latter is unresolved, and it basically appears associated to X1. We
rather argue that X1 is most likely an ULX similar to the
ones identified in other star-forming galaxies given the similarities in the
values of its luminosity and photon index.

\section{Discussion}

\label{discussion}

The analysis of the observational data we have discussed in previous sections
indicates that the nucleus of Haro~2 has experienced very recent star formation,
being dominated by a population of massive stars with ages in the range $3.5-5$
million years and a total mass of $\sim$$4\times 10^6$ \msun. The interstellar
medium seems to be heavily distorted, with a patchy distribution of dust clouds
which blocks most of the UV continuum emitted by a high fraction of these stars,
which are identified only by their contribution to the ionization of the gas and
the heating of the dust particles. Moreover, while Haro~2 is surrounded by a
relatively dense cloud of neutral gas, parts of it have been accelerated by the
release of mechanical energy by these powerful starbursts, opening kinematical
holes which allow the escape of \lyalp\ photons out of the ionized regions,
albeit in the form of P~Cyg profiles. The release of mechanical energy has also
led to powerful, diffuse soft X-ray emission around the central star-forming
knots. Even more, Haro~2 presents a strong hard X-ray emission of the order of
the soft X-ray emission, that is dominated by a single, unresolved source which
is an ULX candidate. Finally, an extended and diffuse component of \lyalp\
emission has been detected whose origin is unclear. This diffuse emission is spatially not
correlated either with the UV continuum, nor even with the Balmer lines
emission or Balmer decrement, but on the other hand it might originate close to the soft X-ray
emitting regions.

\subsection{Past star formation episodes} 

Previous episodes of massive star formation should have contributed to the
chaotic structure of gas and dust in the nucleus of Haro~2. Some authors claim
indeed that star formation has taken place in Haro~2 previously to the current
starbursting episodes.  \citet{Fanelli88} studied signature lines of stars of
different spectral types in spectra taken by the IUE. They found that the
presence of O7-B0 V stars in Haro~2 reveals a young burst in the galaxy having
ocurred within the past $10$ Myr. However, due to the presence of A-type stars,
they also argued that Haro~2 must have experienced two previous starburts, the
most recent having ended less than $20$ Myr ago. From this data,
\citet{Summers01} claimed that the star formation history of Haro~2 is composed
of a recent burst with age $\sim$$5$ Myr, another of $20$ Myr, and a previous one
which took place around $500$ Myr ago. Also, they found that the older the
burst, the larger the number of stars produced, and hence they concluded that
the intensity of the episodes of the star formation must have decreased with
time, maybe due to the lack of gas available for star production.

In order to check the presence of past star-forming episodes we used the models
by \citet{MartinManjon08} which combine codes of evolutionary population
synthesis models, chemical evolution and photoionization to model the intensity
of optical nebular lines, magnitudes and abundances of metals in a star-forming
region. The models assume a history of star formation composed of bursts
equally-separated, attenuated in time and corrected for the efficiency of the
stellar production from the available gas. The production of stars from gas
which has been chemically enriched by stellar winds and supernovae is
consistently considered, as well as the effect of this enrichment on the
intensity of nebular lines produced by photoionization. Our aim was to use the
models to check whether previous star formation episodes are needed to reproduce
the integrated intensity of the emission nebular lines measured by \citet{Moustakas06}.
Lines observed in the spectrum and whose intensities are predicted by
\citet{MartinManjon08} are: \oxiil, O[III]$5007$, O[I]$6300$, \hal,
N[II]$6584$, S[II]$6716$ and S[II]$6731$. Intensity is scaled to \hb\ and
observed lines were corrected for stellar absorption and Galactic and intrinsic
extinction. Since \lha/\lhb$\sim$$2.86$ was assumed to correct for internal
extinction we did not include \hal\ in the analysis to avoid biases towards
models showing similar values.

Given the integrated value measured by \citet{Moustakas06} for Haro~2 \ewhb$=$$44$ \AA, we
only considered models with predicted \ewhb\ in the range $34$ \AA$\leq$\ewhb$\leq 54$
\AA. Models consider bursts equally-separated in time by either $0.05$, $0.1$, $0.5$ and
$1.3$ Gyr. Results show that models which best reproduce the intensity values of the lines
are those 1) having had previous star formation bursts, either each $0.5$ Gyr or $1.3$
Gyr, and 2) last burst having taken place $3-5$ Myr ago. Point 2) agrees with the
ages of the bursts we have calculated, whereas point 1) reinforces the idea by
\citet{Fanelli88} and \citet{Summers01} that intense star formation in Haro~2 took place as much
as hundreds of millions of years before the current episode. Therefore these results
confirm that previous star formation in Haro~2 occurred and is responsible for some of the
features observed in the optical spectrum.

\subsection{X-rays}

Soft X-ray emission in a starburst is expected to be produced by 1) the heating of the
diffuse gas surrounding the stellar clusters up to temperatures of millions of Kelvin, due
to the stellar winds and supernovae injecting mechanical energy into the medium, and 2) by
the emission of supernova remnants (SNRs) during the adiabatic phase. The contribution by
the direct emission of individual stars is negligible when compared to the previous
components \citep{Cervino02b}. Evolutionary population synthesis models CMHK02 predict the
soft X-ray emission of a starburst after defining the physical properties of the burst.
CMHK02 model the component 1) through a Raymond-Smith plasma with $kT=0.5$ keV,
considering that a fraction \xeff\ from the mechanical energy is finally converted into
X-ray luminosity. Typically, values \xeff$=1-10$\% are found for star-forming objects
\citep{Summers01,Summers04,MasHesse08}. On the other hand, the soft X-ray emission by SNRs
is considered to be due to the reversed shock which heats the gas inside the shell and
which is modeled by a composite of three Raymond-Smith plasmas with $kT=0.23$, $0.76$,
and $1.29$ keV. See \citet{Cervino02b} for a more detailed description.

As already explained, using the
{\em ROSAT}/PSPC observation of Haro~2 \citet{Stevens98} obtained values $kT=0.36$ keV for
the temperature of the hot plasma and $N($HI$)=2.2\times10^{21}$ cm$^{-2}$ for the
absorption, which yielded a higher soft luminosity $L_{ROSAT}$$\sim$$1.4\times10^{40}$ \ergs\
when they assumed $Z\sim Z_\odot/10$. \citet{Summers01} argued that the high uncertainties
obtained for $N($HI$)$ by \citet{Stevens98} caused also large uncertainties in the X-ray
luminosity, which was constrained in the range
$L_{ROSAT}$$\sim$$2\times10^{39}-1.4\times10^{42}$ \ergs\ and took
$L_{ROSAT}$$\sim$$2\times10^{39}$ \ergs\ as valid. \citet{Summers01} estimated the injection of
mechanical energy into the medium per unit of time \emech\ by two methods: 1) applying the
model for starburst-driven outflows by \citet{Castor75} and \citet{Weaver77} assuming that
the radius of the superbubble blown by the star-forming activity in Haro~2 is
$R_{\rm{B}}=1.1$ kpc, and 2) using the ratio \emech$/L_{bol}$ from SB99 models, where
$L_{bol}$ was considered to be well represented by the observed far infrared luminosity
\lfir$=1.4\times10^{43}$ \ergs. After performing these calculations for several ages, they
assumed the value \emech$=1.2\times10^{41}$ \ergs\ considering an age of the burst of
$5.8$ Myr. Thus, with their values of \emech\ and $L_{ROSAT}$ \citet{Summers01} obtained
\xeff$\sim$$0.02$ for Haro~2.

We calculated the X-ray luminosity predicted by the CMHK02 models for a composite of
two starbursts with the ages and masses found for Population 1 and Population 2
in Sect. \ref{resuvcont} and included in Table~\ref{poptable}. Metallicity of
the models was fixed to $Z=0.008$ since it is the closest value to the observed
one $Z$$\sim$$Z_\odot/2$ \citep{Davidge89}. As shown in Fig.~\ref{figxspec} soft
X-ray emission is not strictly due to gas heated by the starburst activity (HP
component), but also by the X1 emission (PL component). Actually, as indicated
in Table~\ref{xraylum} soft X-ray emission due to HP is $\sim$$45$\% of the total
soft X-ray emission observed, i.e. gas heated by the starburst activity emitts
$L_{0.4-2.4\rm{\,keV}}$$\sim$$1.1\times10^{39}$ \ergs. CMHK02 can reproduce this
soft X-ray luminosity value for the composite of the bursts explained assuming
\xeff$=0.004$, i.e. the efficiency in the conversion of mechanical energy
injected into the ISM by SNe and stellar winds in Haro~2 into soft X-ray
luminosity is $\sim$$0.4$\%. \citet{Summers01} found the much higher value of
$2$\% since they did not subtract the emission of X1 from the total observed
luminosity. If we consider that the whole total luminosity
$L_{0.4-2.4\rm{\,keV}}$$\sim$$2.4\times10^{39}$ \ergs\ is due to gas heated by the
starburst activity as \citet{Summers01} did, we obtain an efficiency of
$\sim$$1$\% which is in better agreement with their value. When calculating the efficiency \xeff\ for
galaxies where no subtraction of the emission from compact, point-like emitters
is possible (due to lack of spatial resolution, for instance), then comparison
must be made with efficiency values obtained with the total soft X-ray
luminosity. As explained, this is the case for the value reported by
\citet{Summers01}. Given the uncertainties in the calculation of the X-ray
luminosity predicted by the CMHK02 models and in the estimation of the
contribution by X1 to the soft X-ray emission, the range \xeff$=0.004-0.01$
should be assumed for the value of this efficiency in Haro~2.

\subsection{\lyalp\ emission} 

Starbursts in the nucleus of Haro~2 dominate the ultraviolet emission of the
galaxy (mainly by massive stars in dust-free environments), as well as its
ionizing power (by the most reddened young stars), which produces
conspicuous nebular emission lines. The spatial profiles of the UV continuum and
the \lyalp\ and \hal\ emission lines along directions D1 and D2 are shown in
Fig.~\ref{figstispro}, together with the observed ratios \lyalp$/$\hal\ and 
\hal$/$\hb. Also, the profile of the expected \lyalp$/$\hal\ ratio is included,
which was calculated assuming case B recombination with $T_{\rm{e}}=10^{4}$ K
and \ne~$=500$ cm$^{-3}$ (\lyalp$/$\hal~$=8.7$, \hal$/$\hb~$=2.87$), applying
the expected internal dust reddening as derived from the observed \hal$/$\hb\
ratio, as well as the corresponding Galactic extinction. Values of
signal-to-noise ratio larger than $3$ were obtained in the profiles by lowering
the spatial resolution of the emission lines data when needed. Specifically, the
resolution in \hal$/$\hb\ was lowered down to $2.5$ arcsec pixel$^{-1}$, whereas
\lyalp$/$\hal\ was calculated for $0.5$ arcsec pixel$^{-1}$ and $1.0$ arcsec
pixel$^{-1}$ for directions D1 and D2, respectively. As seen in the figure, the
profiles of the UV continuum and the \hal\ emission line are tightly spatially
correlated, as expected since the Balmer lines are mainly produced by the
recombination of gas ionized by the young massive stars.

On the other hand, \lyalp\ is not always spatially coupled to the stellar
continuum, Balmer lines emission or \hal$/$\hb\ ratio. While compact \lyalp\
components \lyai, \lyaiv\ and \lyav\ are located within (or close to) a
Balmer-line-emitting region, diffuse emissions \lyaii\ and \lyaiii\ appear in
regions where the continuum and other lines are very weak. In what follows we
will analyze these 2 sets of regions independently, since their origin is
probably different. \lyai, \lyaiv\ and \lyav\ are apparently produced by
recombination in the gas ionized by the massive stellar clusters, but the
variable conditions of the neutral gas surrounding these regions (mainly
kinematics and column density) determine whether the emission can escape or not,
and whether the line profile is kinematically distorted. The final result is the
alternance of peaks and valleys observed in the \lyalp\ spatial profile between
knots SE and NW along slit D2 (see Fig.~\ref{figstispro}). We want to stress
that this structure is not reflected in the spatial profile of \hal\ since this
line is not affected by resonant scattering.  
 
The \lyalp\ emission identified in regions \lyaii\ and \lyaiii\ shows strikingly different
properties. First of all, the \lyalp$/$\hal\ ratio is much higher in these areas than
close to the stellar clusters (see Fig.~\ref{figstispro}). Ratios as high as
\lyalp$/$\hal~$=3$ and \lyalp$/$\hal~$=6$ are reached along directions D1 and D2,
respectively. These observed ratios are  much larger than predicted applying case B and
the reddening derived from the Balmer decrement, representing \lyalp\ emission $\sim$$3$
and $>$$100$ times higher than expected, respectively. Furthermore, it is remarkable that
along D2, \lyalp$/$\hal\ is highest where \hal$/$\hb\ also reaches its highest value (NW end of the slit),
leading initially to reject the hypothesis that the relatively large \lyalp$/$\hal\ values in the diffuse
regions are originated simply by a low abundance of dust.  

Diffuse \lyalp\ emission similar to \lyaii\ and \lyaiii\ has been found in other local
\lyalp-emitting galaxies. \citet{Hayes05} found that $70$\% of the total \lyalp\ radiation
of the galaxy ESO338-IG04 is diffuse emission located in a halo surrounding the central
regions and lacking strong stellar continuum.  Furthermore, \citet{Hayes07} concluded that
$90$\% of the \lyalp\ emission in Haro~11 is also diffuse and is located in a featureless
halo, where \lyalp\ flux is much higher than expected from recombination, when scaled from
the observed \hal\ intensity. \citet{Atek08} studied local values of the \lyalp$/$\hal\
ratio in a sample of local \lyalp-emitting starburst galaxies. Together with the values
found by \citet{Hayes05} and \citet{Hayes07} for ESO338-IG04 and Haro~11 respectively,
they reported those for IRAS~0833 and NGC~6090. For the latter sources \citet{Atek08} also
found \lyalp\ emission values higher than predicted from case B recombination considering
the \hal\ fluxes observed, when applying the extinction derived from the Balmer
decrement (\hal$/$\hb).

These properties indicate that the origin of the \lyalp\ emission in the diffuse regions
is apparently different from that of the compact knots dominated by the massive clusters.
We postulate that the \lyalp\ emission in these diffuse regions could be enhanced with
respect to \hal\ either by an additional emission component, associated to the hot plasma
identified in the X-ray image, by leaking of the \lyalp\ photons created in the ionized
regions, after multiple scattering within the neutral clouds, by reflection on a dust
screen, or by a combination of the 3 processes. In the next subsections we describe in
some detail these different, non-exclusive models.

\subsubsection{Photoionization by the hot plasma}

For this model we assumed that the \lyalp\ diffuse emissions \lyaii\ and \lyaiii\ in
Haro~2 have been produced by ionization dominated mainly by the hot plasma responsible for
the intense soft X-ray radiation, and not by the massive stellar clusters as for \lyai,
\lyaiv\ and \lyav. As we observe in Fig.~\ref{figx} there seems to be some spatial
correlation between the diffuse \lyalp\ components and the diffuse, soft X-ray emission,
originated by the heating in shock fronts of the interstellar gas surrounding the stellar
clusters. We postulate that this X-ray radiation might have ionized the nebular gas in
these regions, in an environment far from case B recombination conditions.

To check this possibility we used the photionization code {\em CLOUDY} 10.00
\citep{Ferland98}. Ratios \hal$/$\hb\ and \lyalp$/$\hal\ were calculated in two
extreme cases in which a spherical cloud of gas is ionized by a central source,
which is a massive star in Model~1, and a hot plasma in Model~2. In the former
we modeled the continuum with a Kurucz star with $T$$\sim$$4\times10^{4}$ K and
emitting $Q(H)=10^{50}$ s$^{-1}$ ionizing photons, whereas the electronic
density was assumed to be \ne~$=10^{3}$ cm$^{-3}$. Given a typical distance
between the star and the cloud of gas of $\sim$$1$ pc, the value of the
ionization parameter is $U=0.1$. On the other hand, for Model~2 we assumed
a Bremsstrahlung emission with $T=10^{7}$ K and \ne~$=10^{3}$ cm$^{-3}$, keeping
the ionization parameter within $U=10^{-8}-10^{-5}$. This extreme conditions
were chosen to reproduce the physical properties of high-density filaments
compressed by the shock fronts.

The emission line ratios computed for Model~1 are (\hal$/$\hb)$_{\rm{Model\,1}}=2.9$ and
(\lyalp$/$\hal)$_{\rm{Model\,1}}=10.1$, indeed very similar to the predictions of the
standard case B, as expected. On the other hand, for Model~2 we obtained
(\hal$/$\hb)$_{\rm{Model\,2}}$ in the range $3.0-4.0$, and
(\lyalp$/$\hal)$_{\rm{Model\,2}}$ from $7.0$ to $12.0$. It is important to note that Model~2
predicts systematically higher \hal$/$\hb\ values than case B. This implies that if the
contribution by the hot plasma is important, we could overestimate the abundance of dust
if derived assuming case B conditions.

We would expect a physical scenario with a smooth transition from Model~1 to Model~2
conditions as we move away from the vicinity of the massive clusters to more external
regions. Fig.~\ref{figstispro} shows compact \lyalp\ emission, if any, close to the
stellar clusters in Haro~2. Assuming that Model~1 conditions prevail in these regions,
reddening correction as derived from the case B \hal$/$\hb\ ratio should be applied. The
\lyalp$/$\hal\ ratios observed around these regions are generally lower than expected from
case B, even taking into account the effect of reddening. This is especially noticeable
around  \lyaiv\ and \lyav. We know already from UV spectroscopy that scattering in the
outflowing neutral gas surrounding knot SE, together with the relatively high dust
abundance (as derived from the observed \hal$/$\hb\ ratio),  are responsible for the low
escape fraction of the \lyalp\ photons originated on this area. 

On the contrary, the diffuse \lyalp\ emission is originated in regions spatially separated
from the stellar clusters, where the ionization might be dominated by the hot plasma.
Since Model~2 predicts higher intrinsic \hal$/$\hb\ ratios than Model~1, the extinction
derived in the diffuse emitting regions would be much lower than assuming case B, allowing
for higher \lyalp\ escape fractions in these areas. The ratio measured on \lyaii,
\lyalp$/$\hal$<3.0$, is compatible with the results found by \citet{MasHesse03}. They
found that the diffuse \lyalp\ emission exhibits a clear P~Cyg profile all along slit D1,
at any distance from the stellar knot SE. This implies that a significant fraction of the
emitted \lyalp\ photons (the blue wing of the line) has been affected by resonant
scattering, decreasing the outcoming line flux in this area. On region \lyaiii, both the
observed  \hal$/$\hb\  and  \lyalp$/$\hal\ ratios could be consistent with the presence of
shock-heated filaments, with small or none attenuation by dust. Indeed, the strong X-ray
flux in this area would most likely destroy most of the dust particles. The fact that the
\lyalp$/$\hal\ ratio is very close to the intrinsic predictions by Model~2 requires also
that the medium in front of this region is transparent to \lyalp\ photons: either the
column density of neutral gas is small enough, or it is outflowing with a projected
velocity large enough not to affect the \lyalp\ intensity ($v_{out}\ge 400$ km s$^{-1}$).
Unfortunately the spectral resolution is not enough to analyze the
spectral profile of \lyaiii.

Model~2 predicts \lhalx$\sim$$10^{-3}-10^{-2}$, where \lha\ and
$L_{0.4-2.4\rm{\,keV}}$ are the intrinsic \hal\ luminosity produced and the ionizing soft
X-ray luminosity emitted by the hot plasma, respectively. We measured \lhalx\ in both the
regions dominated by the stellar clusters and those showing diffuse \lyalp\ emission. For
the compact regions we found \lhalx$\sim$$300$, while in the diffuse regions, far away
from the stellar clusters, \lhalx$\sim$$1$. This severe decrease in the \lhalx\ ratio when
moving towards regions showing diffuse emission  reinforces the idea of ionization in
these sites being dominated by the hot plasma itself. Values of \lhalx$\sim$$10-100$ are
measured in the intermediate regions, where the ionizing flux of the massive stars
increases with respect to the hot plasma one as we move closer to the stellar clusters.
Although the measured \lhalx\ is always higher than the absolute values predicted by Model~2,
we should be aware that the hot plasma filaments which are expected to ionize the local
surrounding medium can not be resolved, and hence the flux values and ratios obtained are
average values. Measuring \lhalx\ locally would require a higher spatial resolution and
better statistics in the X-ray observation.  However, the observed trend in \lhalx\ does
point towards the diffuse \lyalp\ emission being produced by the ionization caused by the
hot plasma, whereas the compact \lyalp\ emission is consistent with being originated in
the gas ionized by the massive stars.

Finally, while the simulations with {\em CLOUDY} suggest that the hot plasma could dominate the
\lyalp\ emission in the external regions of Haro~2, we have not been able to reproduce all
the spectral features observed in the regions showing \lyaii\ and \lyaiii\ emissions.
Whereas the ionizing hot plasma models, combined with some attenuation,  can reproduce the
observed ratios \lyalp$/$\hal, \hal$/$\hb\ and \lhalx, the predictions for some other
lines (like Fe~II~$4300$ and Fe~II~$6200$) are overestimated when compared with the
observations. Furthermore, other lines present in the spectra such as \oxiil\ and \oxiiid\
are underestimated by these models. Nevertheless, we want to stress that our models are
based on a simplistic scenario assuming spherical geometry, while X-ray-emitting filaments
should be significantly more complex in shape, structure and physical conditions. A more
detailed modelling of the geometry of these filaments is out of the scope of this work. We
want to retain here only the idea that a very hot, X-ray-emitting plasma could generate an
additional \lyalp\ emission component under conditions that would differ significantly
from case B ones, and which might be the major contributor to the diffuse \lyalp\ emission
in Haro~2.

\subsubsection{Scattering by the neutral gas} 

As we have already discussed, \lyalp\ photons are resonantly scattered by neutral hydrogen
atoms. \citet{Verhamme06} have modelled the properties of \lyalp\ radiation transfer
through a neutral medium, assuming different conditions of dust abundance, kinematics of
the gas, column density,...  In the absence of dust, \lyalp\ photons could travel long
distances through the neutral cloud, before leaking from its surface, escaping from the
galaxy. Nevertheless, the multiple scattering would at least modify their distribution in
energy, leading to a two-peak emission profile with almost no emission at the central
wavelength. If the gas has some velocity with respect to the ionized region where the
\lyalp\ photons were produced, a P~Cyg-like profile would be expected. The multiple
scattering increases the mean free path of the photons through the neutral cloud,
increasing the probability of being absorbed by dust. 

The presence of neutral gas around the Haro~2 nuclear starburst is evident from
the observations. The P~Cyg profile of the \lyalp\ emission line along slit D1
indicates that a least knot SE is surrounded by outflowing neutral gas. The
regions where the \lyalp\ emission disappears completely along slit D2 are most
likely also covered by neutral gas clouds, in different kinematical states.
Therefore, it might be possible that the \lyalp\ photons, formed in the ionized
regions close to the stellar clusters, but being scattered by these neutral
clouds, end up leaking at relatively large distances. Since \hal\ would not be
affected by this process, the net result would be an enhancement of the \lyalp\
emission on extended areas far away from the ionized regions, as proposed by
\citet{Hayes07}, which in principle would be consistent with the observations.
However, in that case the Balmer lines would be produced by the ionizing flux
originated in the stellar clusters, and conditions similar to case B would
apply. The observed \hal$/$\hb\ values would indicate a high abundance of dust
particles in front (or mixed with) the clouds surrounding \lyaii\ and \lyaiii\
regions, and they would most likely destroy most of the scattered \lyalp\
photons. Nevertheless, while the \lyalp\ photons would be leaking from the
surface of the neutral clouds, the residual Balmer lines emission in these
regions would probably be originated at a different location, so that the
relatively high extinction derived from the  \hal$/$\hb\ ratio would not apply
to the \lyalp\ emission. A detailed analysis of the \lyalp\ spectral profile of
the \lyaiii\ region, not possible with the low-resolution existing data, would
be required to compare it with the predictions of radiation transfer models.

\subsubsection{Dust reflection} 

If we do not consider the possible contribution to ionization by the hot plasma, and
assume that conditions close to case B apply over most of the nuclear region of Haro~2,
then the high \hal$/$\hb\ ratios (especially on region \lyaiii) would point to the
presence of dense dust clouds or filaments at certain distance from the stellar clusters.
These dust clouds could be reflecting the emission originated close to the starburst
knots. Since UV photons are more efficiently dispersed than optical photons, an
enhancement of the \lyalp\ emission over the Balmer lines in these regions would be
naturally produced, explaining the higher \lyalp$/$\hal\ ratios in these extended regions.
In principle we would expect also \hal$/$\hb$<2.86$ in these regions, but since dispersion
would be so much efficient for \lyalp\ wavelengths than for \hal\ or \hb, it might be well
that the observed \hal\ and \hb\ emissions are dominated by the component originated
locally, and therefore attenuated by the local dust. The local \lyalp\ would be almost
totally destroyed, as expected from the high \ebv\ derived. But the reflected \lyalp\
component (scattered at the surface of the dust clouds, and therefore not affected by the
internal extinction) could be strong enough to produce the observed \lyalp$/$\hal\
values.The main flaw of this model is that it predicts lower \lyalp\ equivalent widths
than observed, since the UV continuum would also be reflected.

\vspace{1em}

These different mechanisms are not mutually exclusive, and each of them could be
contributing to a different extent to the \lyalp\ emission detected in starburst
galaxies like Haro~2. \citet{Scarlata09} proposed indeed a scenario which combines both
the scattering by neutral gas and the reflection by dust clouds, based on a
clumpy distribution of neutral clouds rich in dust, surrounding the ionized
region. While the Balmer lines and the UV continuum would be attenuated by the
dust particles along the line of sight, scattering and/or reflection at the
surface of the neutral clouds would allow the \lyalp\ photons to find their way through
the medium with a rather high efficiency, leading to high \lyalp$/$\hal\ ratios
when observed globally. They argued that this scenario could explain the
integrated values of \lyalp$/$\hal~$=1-5$ measured in a sample of
{\em GALEX} \lyalp-emitters at $z=0.18-0.28$.  Mapping the distribution of the
dust clouds in Haro~2 with very high spatial resolution would be
needed to verify if this scenario could explain the diffuse \lyalp\ emission.

However, we argue that ionization by the hot plasma is the dominating source responsible
for the diffuse emission in Haro~2. Despite the limitations of this model concerning the
predicted intensities of some  emission lines, it can naturally explain the spatial
coupling of the \lyaii\ and \lyaiii\ components with the soft X-ray emission, the measured
values of the \lyalp$/$\hal\ and \lhalx\ ratios and the high escape fraction of \lyalp\
photons from regions with high \hal$/$\hb\ ratios.

If the diffuse \lyalp\ emission detected in Haro~2 and other starburst galaxies is
originated by ionization by hot plasma, and not directly by the massive stars, the use of
\lyalp\ could overestimate the star formation rate in galaxies at hight redshift by a
significant factor (when integrated over the whole galaxy, \citet{Hayes05, Hayes07} found
that the diffuse emission is a dominant fraction of the total \lyalp\ emission observed).
This adds to the difficulty of determining the escape fraction of \lyalp\ photons
originated directly by stellar ionization along the history of the Universe
\citep{Hayes11}. A further correction should be applied to the procedure used to convert
\lyalp\ luminosity into number of ionizing photons, and thus star formation rate.
Calibrating the strength of these corrections  requires a detailed analysis of a larger
sample of \lyalp-emitting starburst galaxies in the local Universe.

\section{Conclusions}
\label{conclusions}

We have carried out a complete multiwavelength spectral and photometric analysis
of the star-forming nucleus of the local \lyalp-emitter Haro~2, focusing on the
\lyalp\ emission and its relation with different starburst-related parameters. UV,
optical, NIR and X-ray images, as well as UV-optical spectral observations with
high spatial and spectral resolution, have been analyzed.

\begin{itemize}

\item We have characterized the starburst of Haro~2 by comparison with
evolutionary population synthesis models, reproducing the UV, \hal\ and FIR
luminosities values as well as the properties of the SiIV and CIV stellar
absorption lines. We have identified a population of young massive stars (ages
$\sim$$3.5-5.0$ Myr) with a total stellar mass $M$$\sim$$4\times10^{6}$ \msun\  (for an
IMF with mass limits of $2-120$ \msun), and whose continuum is affected by a
differential extinction with \ebv\ values in the range \ebv=$0.035 -\geq 0.5$. Stars
located in dust-free regions dominate the UV emission, whereas those within
dust-rich clouds are completely obscured in the UV, but are the main
contributors to the \hal\ and FIR luminosities of the galaxy.

\item We have identified a diffuse soft X-ray emission extended over the whole
nucleus. This component is attributed to the gas heated by the release of
mechanical energy  by the stellar winds and supernovae from the starburst, as
well as to supernova remnants. We have estimated that $\sim$$0.4-1$\% of the
mechanical energy injected is converted into soft X-ray emission, which is a
typical value for this kind of sources. A hard, unresolved X-ray emission is
located on the SE star-forming knot, apparently originated by an
UltraLuminous X-ray (ULX) source related to the starburst episode.

\item Both compact and diffuse \lyalp\ emission components, together with regions of total
absorption, are observed along the major and minor axes of the star-forming nucleus of
Haro~2. Intensities of the compact \lyalp\ emission are lower than those expected from
\hal\ fluxes and case B recombination theory, considering dust extinction. On the other
hand, the diffuse \lyalp\ emission is stronger than expected from \hal, and extends over
$850$ pc and $>600$ pc along minor and major axes, respectively. We have found that the
\lyalp\ emission is spatially decoupled from the UV continuum, the Balmer lines emission
and the \hal$/$\hb\ ratio, but the diffuse \lyalp\ component is apparently correlated with
the diffuse soft X-ray emission.

\item Outflowing of the neutral hydrogen surrounding the stellar clusters,
energized by the starburst activity, enables \lyalp\ photons from both compact
and diffuse components to escape from Haro~2, leaving the signature of a clear
P~Cyg profile in the emission line. Nevertheless, the irregular distribution and
kinematical properties of the expanding neutral gas lead the \lyalp\ photons
produced in recombination sites to escape only in certain regions. \hal\ photons
are not affected at all by the kinematics of the neutral gas, and escape
directly from all ionized regions. This explains the spatial decoupling between
the \lyalp\ and \hal\ compact emissions.

\item Compact \lyalp\ emission is produced by recombination in the gas ionized by the
massive stellar clusters. The combined effect of resonant scattering by neutral gas and
attenuation by dust, as traced by the P~Cyg profiles and the high values of \hal$/$\hb\
ratio respectively, decreases the escape fraction of the \lyalp\ photons and yields 
observed fluxes below the case B recombination predictions.

\item On the other hand, the diffuse \lyalp\ emission has apparently a rather
different origin. We argue that the shock fronts heated to X-ray emitting
temperatures by the release of mechanical energy could ionize the surrounding
nebular gas under conditions far from case B recombination, as suggested by the
spatial correlation between the soft X-ray emission and the diffuse \lyalp\
component. Simple modelling with {\em CLOUDY} reproduces indeed the \lyalp$/$\hal,
\hal$/$\hb\ and \lhalx\ ratios measured in this region.

\item Nevertheless, our results do not reject other physical processes proposed previously
that could also contribute to enhance the intensity of the diffuse \lyalp\ emission line.
\lyalp\ photons, though originated close to the massive stars, could be leaking from the surface of
distant neutral clouds after multiple scattering within the gas, depending on its
distribution and the abundance of dust. Moreover, under certain circumstances the dust
clouds could even contribute to enhance the diffuse \lyalp\ emission by differential
reflection.   

\end{itemize}

\begin{acknowledgements}

HOF and JMMH are partially funded by Spanish MICINN grants CSD2006-00070 ({\em
CONSOLIDER GTC}), AYA2010-21887-C04-02 ({\em ESTALLIDOS}) and AYA2008-03467/ESP.
HOF is funded by Spanish FPI grant BES-2006-13489. MH received support from the
Agence Nationale de la Recherche (ANR-09-BLAN-0234-01). HA and DK are supported
by the Centre National d'\'Etudes Spatiales (CNES) and the Programme National de
Cosmologie et Galaxies (PNCG). We want to acknowledge the use of the {\em
Starburst99} models and the NASA/IPAC Extragalactic Database (NED). We are very
grateful to the {\em ESTALLIDOS} collaboration for its scientific support. Also,
we thank Giovanni Miniutti for his useful comments on X-ray modelling of HMXBs.
We are very grateful to Herv\'e Bouy for his help in creating
Figs.~\ref{figharo},~\ref{fignicwf} and~\ref{fignicwfx} and to Nuria Hu\'elamo
and Herv\'e Bouy for fruitful discussions about continuum subtraction and PSF
issues. We also thank Ricardo Amor\'in for allowing us to show the {\em
NOT}/ALFOSC image of Haro~2. We want to thank Mercerdes Moll\'a and Mariluz
Mart\'in-Manj\'on for their help on the use and interpretation of their models.
We are also grateful to Valentina Luridiana and Sebastiano Cantalupo for giving
us figures, data and insight on collisional excitation and recombination theory.
This paper was based on observations with {\em Hubble Space Telescope} and {\em
Chandra}.

\end{acknowledgements}

\end{document}